\documentclass[a4paper,11pt]{article}
\pdfoutput=1 

\usepackage{jcappub} 

\usepackage[T1]{fontenc} 

\usepackage{mn2e_journal_style}

\newcommand{\thetas}{\theta_{\rm s}}
\newcommand{\Planck}{{\em Planck}}
\newcommand{\Euclid}{{\em Euclid}}
\newcommand{\WFIRST}{{\em WFIRST}}
\newcommand{\eROSITA}{{\em eROSITA}}
\newcommand{\COBE}{{\em COBE}}
\newcommand{\CORE}{{\em CORE}}
\newcommand{\sigt}{\sigma_{\thetas}}
\newcommand{\Ft}{F_{\thetas}}
\newcommand{\jnu}{j_\nu}		
\newcommand{\Ts}{T_{\thetas}}

\newcommand{\lsim}{\raise0.3ex\hbox{$<$\kern-0.75em\raise-1.1ex\hbox{$\sim$}}}
\newcommand{\gsim}{\raise0.3ex\hbox{$>$\kern-0.75em\raise-1.1ex\hbox{$\sim$}}}

\def\1{\'\i }
\newcommand{\corep}{{\em CORE}}
\newcommand{\vv}{{\mathbf v}}
\newcommand{\vx}{{\mathbf x}}
\newcommand{\vn}{\hat{\mathbf{n}}}
\newcommand{\vrv}{{\mathbf r}}
\newcommand{\covC}{{\mathbf C}}

\title{Exploring Cosmic Origins with \CORE: Cluster Science}

\author[1]{J.-B. Melin,\note{Corresponding author.}}
\author[2,3]{A. Bonaldi,}
\author[2]{M. Remazeilles,}
\author[4,5]{S. Hagstotz,}
\author[6]{J.M. Diego,}
\author[7]{C. Hern{\'a}ndez-Monteagudo,}
\author[8,9]{R.T.~G{\'e}nova-Santos,}
\author[10]{G. Luzzi,}
\author[11]{C.J.A.P.~Martins,}
\author[12,5]{S.~Grandis,}
% ...

\affiliation[1]{\scriptsize CEA Saclay, DRF/Irfu/SPP, 91191 Gif-sur-Yvette Cedex, France}
\affiliation[2]{\scriptsize Jodrell Bank Centre for Astrophysics, Alan Turing Building, School of Physics and Astronomy, The University of Manchester, Oxford Road, Manchester, M13 9PL, U.K.}
\affiliation[3]{\scriptsize SKA Organization, Lower Withington Macclesfield, Cheshire SK11 9DL, U.K.}
\affiliation[4]{\scriptsize Universit\"ats-Sternwarte, Fakult\"at f\"ur Physik, Ludwig-Maximilians-Universit\"at M\"unchen, Scheinerstr. 1, 81679 M\"unchen, Germany}
\affiliation[5]{\scriptsize Excellence Cluster Universe, Boltzmannstr. 2, 85748 Garching, Germany}
\affiliation[6]{\scriptsize IFCA, Instituto de F{\'i}sica de Cantabria (UC-CSIC), Av. de Los Castros s/n, 39005 Santander, Spain}
\affiliation[7]{\scriptsize Centro de Estudios de F{\'i}sica del Cosmos de Arag{\'o}n (CEFCA), Plaza San Juan, 1, planta 2, E-44001 Teruel, Spain}
\affiliation[8]{\scriptsize Instituto de Astrof{\'i}sica de Canarias, C/V{\'i}a L{\'a}ctea s/n, La Laguna, Tenerife, Spain}
\affiliation[9]{\scriptsize Departamento de Astrof{\'i}sica, Universidad de La Laguna (ULL), La Laguna, Tenerife, 38206 Spain}
\affiliation[10]{\scriptsize Dept. of Physics, Sapienza, University of Rome, Piazzale Aldo Moro, Rome, I-00185 Italy}
\affiliation[11]{\scriptsize Centro de Astrof\'{\i}sica da Universidade do Porto and IA-Porto, Rua das Estrelas, 4150-762 Porto, Portugal}
\affiliation[12]{\scriptsize Faculty of Physics, Ludwig-Maximilians-Universit\"at, Scheinerstr. 1, 81679 Munich, Germany}
% ...

\emailAdd{jean-baptiste.melin@cea.fr}

\abstract{We examine the cosmological constraints that can be achieved with a galaxy cluster survey with the future \CORE\ space mission. Using realistic simulations of the millimeter sky, produced with the latest version of the \Planck\ Sky Model, we characterize the \CORE\ cluster catalogues as a function of the main mission performance parameters.  We pay particular attention to telescope size, key to improved angular resolution, and discuss the comparison and the complementarity of \CORE\ with ambitious future ground-based CMB experiments that could be deployed in the next decade.\\
A possible \CORE\ mission concept with a 150\,cm diameter primary mirror can detect of the order of 50,000 clusters through the thermal Sunyaev-Zeldovich effect (SZE).  The total yield increases (decreases) by 25\% when increasing  (decreasing) the mirror diameter by 30\,cm. The 150\,cm telescope configuration will detect the most massive clusters ($>10^{14}\, M_\odot$) at redshift $z>1.5$ over the whole sky, although the exact number above this redshift is tied to the uncertain evolution of the cluster SZE flux-mass relation; assuming self-similar evolution, \CORE\ will detect $\sim 500$ clusters at redshift $z>1.5$.  This changes to 800 (200) when increasing (decreasing) the mirror size by 30\,cm.  \CORE\ will be able to measure individual cluster halo masses through lensing of the cosmic microwave background anisotropies with a 1-$\sigma$ sensitivity of $4\times10^{14} M_\odot$, for a 120\,cm aperture telescope, and $10^{14} M_\odot$ for a 180\,cm one. \\
From the ground, we estimate that, for example, a survey with about 150,000 detectors at the focus of 350\,cm telescopes observing 65\% of the sky from Atacama would be shallower than \CORE\ and detect about 11,000 clusters, while a survey from the South Pole with the same number of detectors observing 25\% of sky with a 10\,m telescope is expected to be deeper and to detect about 70,000 clusters. When combined with such a South Pole survey, \CORE\ would reach a limiting mass of  $M_{500} \sim 2-3 \times 10^{13} M_\odot$ and detect 220,000 clusters (5 sigma detection limit).\\
Cosmological constraints from \CORE\ cluster counts alone are competitive with other scheduled large scale structure surveys in the 2020's for measuring the dark energy equation-of-state parameters $w_0$ and $w_a$ ($\sigma_{w_0}=0.28$, $\sigma_{w_a}=0.31$).  In combination with primary CMB constraints, \CORE\ cluster counts can further reduce these error bars on $w_0$ and $w_a$ to 0.05 and 0.13 respectively, and constrain the sum of the neutrino masses, $\Sigma m_\nu$, to $39\,{\rm meV}$ (1 sigma).\\
The wide frequency coverage of \CORE, 60 - 600\,GHz, will enable measurement of the relativistic thermal SZE by stacking clusters.  Contamination by dust emission from the clusters, however, makes constraining the temperature of the intracluster medium difficult. The kinetic SZE pairwise momentum will be extracted with $S/N=70$ in the foreground-cleaned CMB map. Measurements of $T_{\rm CMB}(z)$ using \CORE\ clusters will establish competitive constraints on the evolution of the CMB temperature:  $(1+z)^{1-\beta}$, with an uncertainty of $\sigma_\beta \lesssim 2.7\times 10^{-3}$ at low redshift ($z \lesssim 1$).
The wide frequency coverage also enables clean extraction of a map of the diffuse SZE signal over the sky, substantially reducing contamination by foregrounds compared to the \Planck\ SZE map extraction.  Our analysis of the one-dimensional distribution of Compton-$y$ values in the simulated map finds an order of magnitude improvement in constraints on $\sigma_8$ over the \Planck\ result, demonstrating the potential of this cosmological probe with \CORE.
}

\begin{document}
\maketitle
\flushbottom

\section{Introduction}
\label{sec:intro}
Galaxy clusters are important cosmological probes, primarily because their abundance and their evolution are very sensitive to the growth rate of large-scale structure \citep{wang98,allen2011}.  This makes them powerful tools for constraining dark energy and possible modifications to gravity \citep{weinberg2013}, and motivates large cluster surveys \citep{haiman01}.  Clusters can be selected as overdensities of galaxies observed in the visible and/or near-infrared (NIR) bands 
\citep[e.g.,][]{gladders00,redmapperdes2016}, as extended sources of X-ray emission \citep[e.g.,][]{rosati2002} and  through the Sunyaev-Zel'dovich effect \citep[SZE,][]{sz1972} \citep{staniszewski09,hasselfield2013,planck2013xxix,bleem2015}.  Surveys in all these wavebands will produce catalogs containing tens of thousands of clusters in the coming decade.  These missions include stage-IV dark energy observatories such as the \Euclid~\citep{laureijs11} and the \WFIRST~\citep{wfirst2015} space missions, the Large Synoptic Survey Telescope \citep[LSST,][]{LSST09}, the \eROSITA\ X-ray satellite \citep{merloni12}, and the next generation of cosmic microwave background (CMB) experiments, such as the Advanced Atacama Cosmology Telescope \citep[AdvACT,][]{niemack2010}, the South Pole Telescope with third generation detector technology \citep[SPT-3G,][]{benson2013} and the proposed CMB-S4 \citep{cmbs4-2016}.

In this paper, we study cluster science that would be enabled by the \CORE\ mission. The mission is proposed to survey the sky in intensity and polarization across 19 broad-bands spanning the frequency domain from 60 to 600\,GHz.  It is proposed in response to the European Space Agency's (ESA) call for a medium-class mission for its M5 opportunity. This paper is one of a series presenting the \CORE\ science goals.  

\CORE\ will image galaxy clusters through their SZE, in which the Cosmic Microwave Background photons undergo inverse Compton scattering off electrons from the hot intracluster gas, leaving a cold spot in the CMB at frequencies below 220~GHz and a hot spot above this frequency. \CORE\ data will enable the construction of a large catalogue of clusters detected via the thermal SZE (tSZE) out to redshifts $z>1.5$ and their use to constrain cosmological parameters, including the dark energy equation-of-state.  The advantage of tSZE detection with respect to other cluster detection techniques is the excellent control it offers on the survey selection function, which is crucial for cosmological applications.  

Another critical aspect of this research program is the calibration of the mass-observable scaling relation.  This is a difficult task, because it requires cluster mass measurements.  Systematic uncertainties in the scaling relations currently limit cluster constraints on cosmological parameters~\citep{planck2014-a30,dehaan16}.   Gravitational lensing provides the most robust mass measurements~\citep{vonderlinden2014,hoekstra2015,umetsu2016}, and the large optical/NIR imaging surveys (e.g., \Euclid, \WFIRST\ and LSST) will use gravitational lensing shear observations to calibrate optical/NIR-mass scaling relations for their surveys.  

For its part, \CORE\ will be able to calibrate the tSZE signal-mass scaling relation through lensing of the cosmic microwave background (CMB) anisotropies in temperature and polarization~\citep{melin2015,baxter2015,madhavacheril2015}.  This CMB-lensing methodology makes \CORE\ self-sufficient for cosmology with cluster counts; moreover, the method enables mass measurements to higher redshifts than possible with galaxy (shear) lensing, which will be important for clusters at $z>1$.  This is important not only for the \CORE\ cluster sample, but will be valuable for cluster surveys proposed by \Euclid, \WFIRST, LSST and \eROSITA.
 
The broad frequency coverage of \CORE\ opens the door to additional cosmological studies with clusters. It will be possible to measure the relativistic corrections (rSZE) in a large sample of massive clusters, which then provide direct measurements of cluster temperatures.  The kinetic SZE (kSZE) effect will be used to measure pairwise peculiar velocities of clusters, thereby probing the instantaneous rate of structure growth and hence constraining modifications of General Relativity.  Finally, the tSZE can be used for an accurate test of a basic tenant of the standard cosmological model: the redshift dependence of the CMB temperature.  

We quantify the scientific reach of \CORE\ in each of these areas using detailed simulations of the sky and mission performance.  We consider what \CORE\ can achieve alone, comparing to a ground-based survey representative of what can be achieved with a future CMB-S4 observatory, as well as the added value of combining the ground-based observatory and space mission data sets.  Particular attention is given to the impact of the choice of the \CORE\ primary telescope aperture (120, 150 and 180\,cm) in each case.

The paper is organized as follows. In Section~\ref{sec:skymaps}, we describe our simulations, followed by a discussion of the expected \CORE\ cluster catalogues in Section~\ref{sec:clustercats}. Section~\ref{sec:analyses} presents forecasts for a variety of studies using the \CORE\ sample. This section includes Subsection~\ref{sec:clustercounts}, which describes the expected constraints on cosmological parameters from the cluster counts that depend on the precision with which we can calibrate the tSZE signal-mass scaling relation with CMB lensing, a topic that is presented in Subsection~\ref{sec:clustermass}.  In Subsection~\ref{sec:relativistic}, we forecast the potential of using \CORE\ to measure the intracluster medium temperature through the relativistic SZE.  Subsection~\ref{sec:kSZE} contains a discussion of the kinetic SZE and a forecast of associated cosmological constraints. In Subsection~\ref{sec:cmb_temp}, we consider how well we will be able to constrain the redshift evolution of the CMB temperature.  We then turn briefly in Section~\ref{sec:diffuse} to science related to the SZ map extracted from the \CORE\ frequency maps.  We conclude in Section~\ref{sec:conclusions}.

Throughout, we adopt the \Planck\ 2015 $\Lambda$CDM best-fit cosmological parameters~\citep[Table 9 of][]{planck2015i}: $h=0.678$, $\Omega_{\rm m}=1-\Omega_\Lambda=0.308$, $\Omega_{\rm b}=0.0484$, $n_s=0.9677$, and $\sigma_8=0.815$.

\section{Synthetic sky maps}

\label{sec:skymaps}

We create synthetic observations using the current version of the \Planck\ Sky Model~\citep{delabrouille2013}. The maps contain primary CMB, galactic emission (dust, free-free and synchrotron), cosmic infrared background, radio and infrared point sources, and cluster signal. 
We pay careful attention to the cluster signal. The clusters are simulated using the Delabrouille-Melin-Bartlett model~\citep{delabrouille2002}: clusters are drawn from the Tinker mass function \citep{Tinker2008} and their intracluster medium pressure is modeled using the circular generalized NFW profile~\citep{nagai2007,arnaud2010}. The clusters are then placed at random sky positions. We assume clusters are isothermal~\citep[adopting the M-T relation from][]{pointecouteau2005}, derive the density profile from the pressure, and model the non-relativistic and relativistic thermal Sunyaev-Zel'dovich effects~\citep{itoh1998,sazonov1998,challinor1998}. We also include the kinetic SZE, assuming uncorrelated Gaussian velocities with zero mean and a standard deviation extracted using linear theory.  In addition to the tSZE and the kSZE, we include emission from dust within clusters, which is an improvement on previous simulations of this kind.  We adopt a modified blackbody spectrum ($\beta=1.5$, $T_d=19.2{\rm K}$) for the dust in clusters~\citep{comis2016}, and we use a spatial profile that is more extended than that of the pressure that has been found to be a good fit to stacked \Planck\ clusters~\cite{melin2017}.

\begin{table}[tbp]
\centering
\begin{tabular}{|c|c|c|}
\hline
Channel & Beam FWHM & Noise $\Delta T$ \\
$\left [ {\rm GHz} \right ] $& [arcmin] & [$\mu{\rm K}$-${\rm arcmin}$] \\
\hline
 60 & 14.3 & 5.3 \\
 70 & 12.3 & 5.0 \\ 
 80 & 10.8 & 4.8 \\ 
 90 & 9.7 & 3.6 \\
100 & 8.7 &  3.5 \\
115 & 7.6 &  3.5 \\
130 & 6.8 &  2.8 \\
145 & 6.1 &  2.5 \\
160 & 5.6 &  2.6 \\
175 & 5.2 &  2.5 \\
195 & 4.7 &  2.5 \\
220 & 4.2 &  2.7 \\
255 & 3.7 &   4.0 \\
295 & 3.2 &  5.2 \\
340 & 2.8 & 7.8 \\
390 & 2.4 & 15.6 \\
450 & 2.1 & 32.5 \\
520 & 1.8 & 82.4 \\
600 & 1.6 & 253.4 \\
\hline
\end{tabular}
\caption{\label{tab:core} Central frequencies of the \CORE\ observing bands with the associated angular resolution and expected noise level for a 150\,cm telescope. For simplicity, we have assumed the frequency bands are Dirac $\delta$ functions.}
\end{table}

We observe this synthetic sky using five instruments: three versions of the proposed space mission, \CORE-150, \CORE-120, \CORE-180, and two possible components of a future ground-based observatory that we call CMB-S4 (Atacama) and CMB-S4 (South Pole), which are representative of what one could think of building in the next decade. 

\CORE\ as proposed in answer to the M5 call of ESA is not fully optimized for SZ science, its main focus and design driver being CMB polarisation. In this paper we consider as a baseline study case a modest extension of the mission concept proposed to M5, \CORE-150, better suited to SZ science.  \CORE-150 has the same frequency channels as the mission proposed to M5, but a slightly larger telescope (150\,cm diameter  aperture) and slightly better sensitivity (an improvement by a factor of $\sqrt{2}$ in sensitivity, which could be obtained straightforwardly with either dual polarisation detectors, or a mission duration extended by a factor of 2). The angular
resolution and sensitivity of each frequency channel of the instrument is given in Table~\ref{tab:core}. We also consider two other scenarios with a smaller (120\,cm) or larger (180\,cm) telescope, but same frequency channels and same raw sensitivity per channel. Only the size of the instrument beam is scaled, for each channel, by a factor of 150/120 or 150/180.

\begin{table}[tbp]
\centering
\begin{tabular}{|c|c|c|}
\hline
Channel & Beam FWHM & Noise $\Delta T$ \\
$\left [ {\rm GHz} \right ] $& [arcmin] & [$\mu{\rm K}$-${\rm arcmin}$] \\
\hline
 40 & 12.4 &  5.3 \\
 95 & 5.2 &  1.5 \\
150 & 3.5 &  1.5 \\
220 & 2.4 &  4.3  \\
270 & 2.0 & 8.5  \\
\hline
\end{tabular}
\caption{\label{tab:atacama} Central frequencies of the CMB-S4 (Atacama) observing bands with the associated resolution and noise level. For simplicity, we have assumed the bands are Dirac delta functions.}
\end{table}

\begin{table}[tbp]
\centering
\begin{tabular}{|c|c|c|}
\hline
Channel & Beam FWHM & Noise $\Delta T$ \\
$\left [ {\rm GHz} \right ] $& [arcmin] & [$\mu{\rm K}$-${\rm arcmin}$] \\
\hline
 40 & 3.8 &  3.2 \\
 95 & 1.6 &  0.9 \\
150 & 1.1 &   0.9 \\
220 & 1.0 &  2.7 \\
270 & 0.9 &  5.3 \\
\hline
\end{tabular}
\caption{\label{tab:southpole} Central frequencies of the CMB-S4 (South Pole) observing bands with the associated resolution and noise level.  For simplicity, we have assumed the bands are Dirac delta functions.}
\end{table}

We consider two study cases for the ground-based observatory, oberving from either the South Pole or the Atacama plateau. We assume that the South Pole site will be equipped with an antenna comparable to the current 10\,m SPT~\cite{carlstrom11}, but a large focal plane array of 155,000 detectors observing the sky in five different frequency channels, and that the Atacama site will be equipped with several new 3.5\,m telescopes (similar to those used by POLARBEAR~\cite{kermish2012}), with the same detector count observing in the same frequency bands, between 40 and 270~GHz. Observing 65\%/25\% of the sky from the Atacama plateau and the South Pole, respectively, during two years of effective observations (i.e., assuming 100\% efficiency) leads to the sensitivities given in Tables~\ref{tab:atacama} and \ref{tab:southpole}. We note that effective observing efficiencies are typically of the order of 20\% rather than 100\%, and hence that the actual observations would require significantly more time (about 10 years) to complete.

\begin{figure}[ptb]
\centering
\includegraphics[width=.67\textwidth]{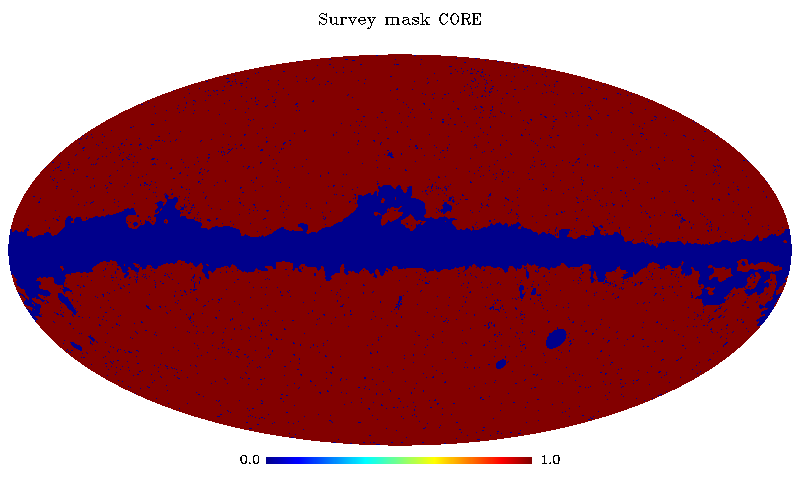}\\
\includegraphics[width=.67\textwidth]{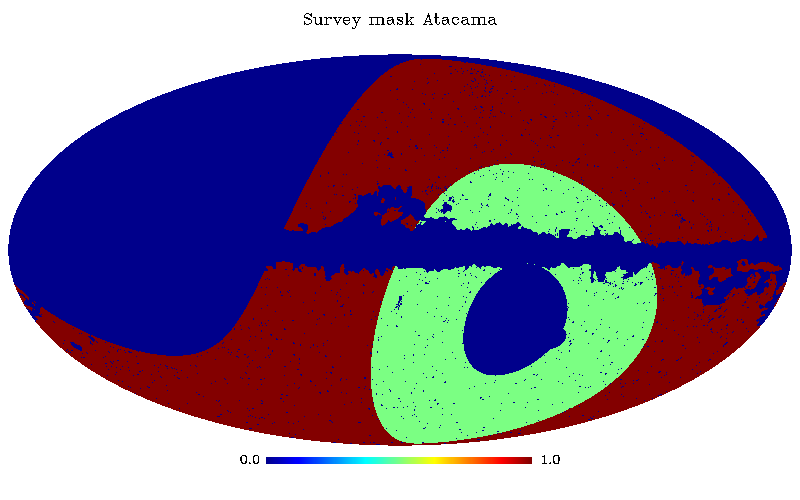}\\
\includegraphics[width=.67\textwidth]{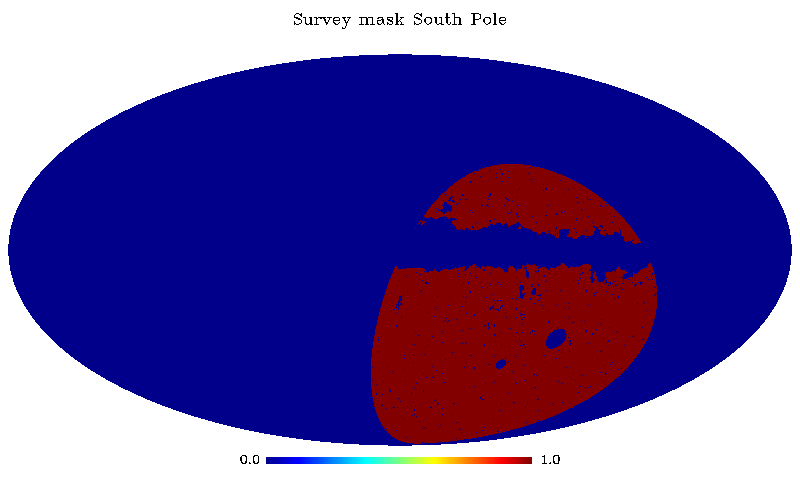}
\caption{\label{fig:mask} Survey masks in galactic coordinates for the \CORE\ mission ($\sim85\%$ of the sky, top), CMB-S4 (Atacama, middle), where the red region ($\sim38\%$ of the sky) is observed from Atacama only, and the green region ($\sim17\%$ of the sky) from both Atacama and the South Pole, and CMB-S4 (South Pole, bottom). }
\end{figure}

None of the five considered experiments -- \CORE-150, \CORE-120, \CORE-180, CMB-S4 (Atacama), CMB-S4 (South Pole) -- will use the full sky for cluster science because of the galactic contamination and -- in the case of ground-based facilities -- the limited sky accessibility. Given the current mission concept, \CORE\ will be able to use $\sim 85\%$ of the sky.  We use in our present analysis the \Planck\ survey mask built for the cluster catalogue (see Fig.~\ref{fig:mask}, top panel). The Atacama/South Pole mask corresponds to observations with a maximum zenith angle of 45/60~deg around the corresponding latitudes of each site -22.9/-90~deg respectively. We use for the Atacama/South Pole surveys the same galactic mask as for the space mission, which masks regions most contaminated by the Milky Way (see two bottom panels of Fig.~\ref{fig:mask}).

\section{Cluster Catalogues}
\label{sec:clustercats}

In our analysis, we follow the procedure adopted by the Planck collaboration for cluster science, which are described in detail elsewhere~\cite{planck2011viii,planck2013xxix,planck2016szcat}. For each of the five surveys -- \CORE-120, \CORE-150, \CORE-180, CMB-S4 (Atacama), CMB-S4 (South Pole) -- we divide the all-sky maps into 504 overlapping tangential patches of $10 \times 10 \deg^2$. For each patch, we compute the noise power matrix $\vec{P}(\vec{k})$ corresponding to instrumental noise and sources of astrophysical signal except for the tSZE. We then estimate the expected noise level $\sigt$ through a Matched Multifilter optimized for tSZE detection~\citep{melin2006}
\begin{eqnarray}
\label{eq:sigt}
\sigt          & \equiv & \left[\int d^2k\; 
     \vec{\Ft}^t(\vec{k})\cdot \vec{P}^{-1}(\vec{k}) \cdot
     \vec{\Ft}(\vec{k})\right]^{-1/2}
\end{eqnarray}
with $\vec{\Ft}(\vec{k}) \equiv \vec{\jnu} \Ts(\vec{k})$ the column vector containing the beam convolved cluster profile at each frequency $\Ts(\vec{k})$ and the expected frequency dependance of the tSZE $\vec{\jnu}$.
We then re-assemble the 504 $\sigt$ functions into a single all-sky HEALPix map. We thus obtain an all-sky tSZE noise map $\sigt(l,b)$ for each experiment.
Clusters from the simulated catalogue are considered as being detected by a given experiment with $S/N \geqslant 5$ if their integrated tSZE flux $Y$ and size $\thetas$ obey
\begin{equation}
Y \geqslant 5 \sigt(l,b) ,
\end{equation}
and if they are located inside the survey area of the experiment. Note that this selection criterion is similar to the selection applied to build the \Planck\ catalogues, and that we carefully normalize the tSZE flux-mass relations in our simulations so our predicted counts are compatible with the cluster counts observed by \Planck.

Expected counts are shown in Table~\ref{tab:counts}. \CORE-150 will detect of the order of 50,000 clusters over 85\% of sky while CMB-S4 (South Pole) will detect $\sim 70,000$ clusters over 25\% of sky. CMB-S4 (Atacama) is expected to be shallower with $\sim 7,000$ clusters over the 38\% sky that do not overlap with the South Pole survey, and about 11,000 clusters in total over its 55\% useful sky.  Combining the multi-frequency \CORE-150 and higher angular resolution CMB-S4 (South Pole) datasets would enable significant reduction in the effect of astrophysical noise sources, such as galactic dust or infrared point sources, allowing one to lower the mass limit significantly and to detect around $200,000$ objects in the 25\% sky visible from the South Pole. This large increase in cluster counts is possible thanks to the large number of \CORE\ observing bands at and between CMB-S4 (South Pole) frequencies, which allow for a more efficient reduction of foreground contamination. Changing the \CORE\ telescope size to 120cm/180cm would lead to a loss/gain of $\sim 25\%$ in the number of detected clusters.

\begin{figure}[ptb]
\centering
\includegraphics[width=.5\textwidth]{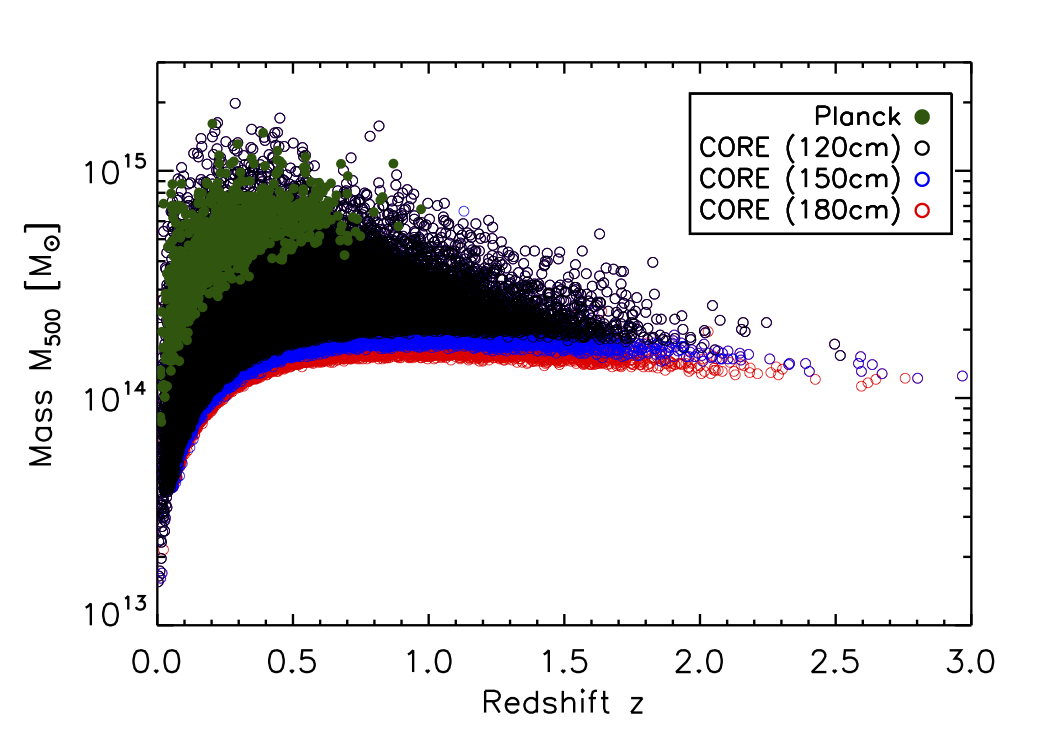}\\
\includegraphics[width=.5\textwidth]{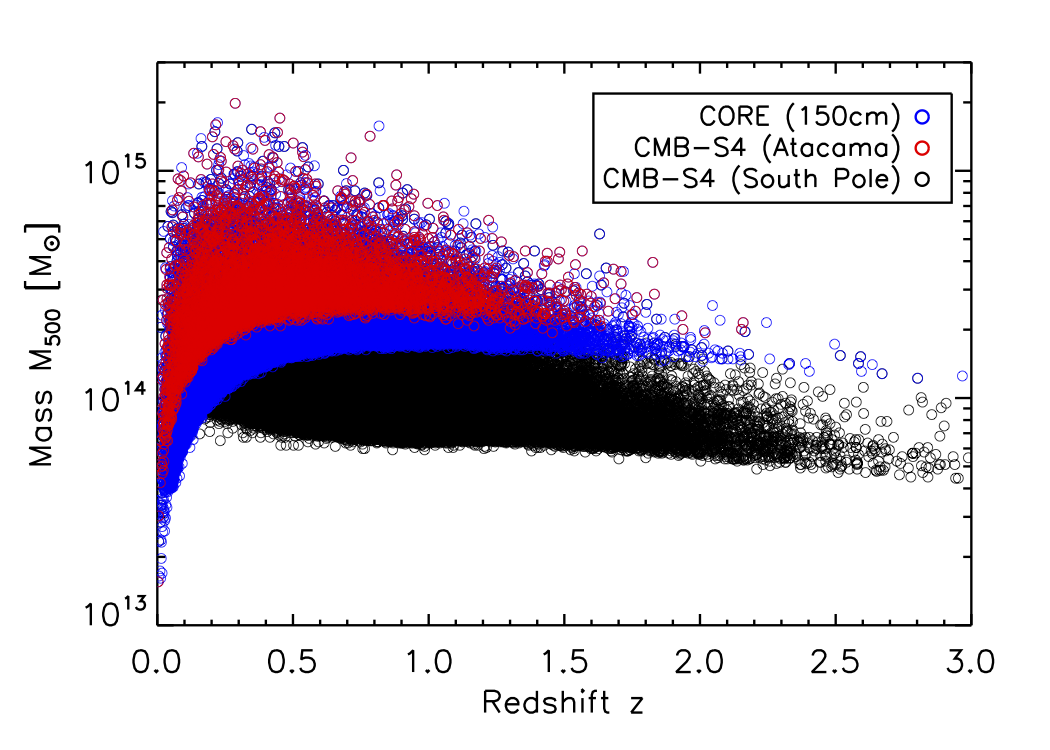}\\
\includegraphics[width=.5\textwidth]{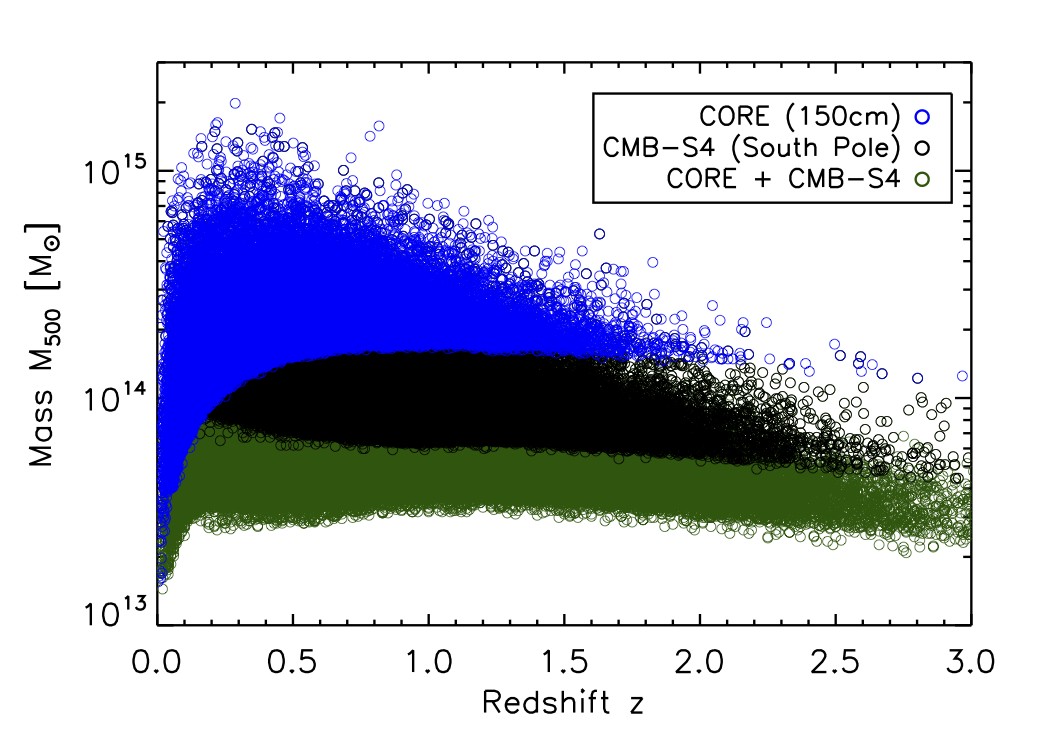}
\caption{\label{fig:mz} Expected mass - redshift distribution for \CORE\ and CMB-S4 cluster samples. {\it Top:} for the three considered \CORE\ apertures. {\it Middle:} for our fiducial \CORE\ aperture and the two CMB-S4 sites. {\it Bottom:} for our fiducial \CORE\ aperture size, CMB-S4 at South Pole and for a joint CMB-S4 and \CORE\ cluster extraction over the South Pole sky.}
\end{figure}

Fig.~\ref{fig:mz} presents the expected mass $M_{500}$ - redshift $z$ distribution of the detected clusters. \CORE\ will detect clusters down to a limiting mass $M_{500}$ lying between $10^{14} M_\odot$ and $2 \times 10^{14} M_\odot$, while CMB-S4 (Atacama) is shallower with a limiting mass between $2 \times 10^{14} M_\odot$ and $3 \times 10^{14} M_\odot$. CMB-S4 (South Pole) is deeper and should reach masses approaching $5 \times 10^{13} M_\odot$. The combination of \CORE-150 and CMB-S4 would permit us to reduce the mass threshold to between $2 \times 10^{13} M_\odot$ and $3 \times 10^{13} M_\odot$. The combination of \CORE\ and CMB-S4 (South Pole) should allow -- for the first time -- the possibility of tSZE cluster selection in the redshift range $2<z<3$.  The exact number is rather uncertain, because it depends on the evolution of the intracluster medium properties in redshift ranges that have not yet been sampled. Moreover, predictions from hydrodynamical simulations have not yet reached a consensus on the tSZE properties of such high redshift clusters~\citep{battaglia2012,kay2012,sembolini2014,lebrun2016,truong2016,barnes2016}.  In our synthetic observations, we have assumed that the local scaling law $Y-M_{500}$ evolves self-similarly over the full redshift range.  If this is the case, \CORE-150 should detect $\sim 500$ objects at $z>1.5$, CMB-S4 would find around $\sim 5,000$ and a combination of the two would allow one to increase this number by as much as a factor of four to $\sim 20,000$.  The number of expected high $z$ clusters for each survey is provided in the right column of Table~\ref{tab:counts}.\\

In comparison, the \eROSITA\ mission (launch in 2017) is expected to detect clusters up to $z=1.5$ in the X-ray, and the \Euclid\ mission (launch in 2020) will reach $z=2$ in optical/NIR. \CORE\ will complement these two large X-ray and optical cluster experiments in the millimeter range and will enable the detection of many high redshift clusters.

\begin{table}[tbp]
\centering
\begin{tabular}{|c|c|c|c|}
\hline
Experiment & $N_{\rm clus}$ & $N_{\rm clus}/{\rm deg}^2$ & $N_{\rm clus}(z>1.5)$ \\
\hline
\CORE-120 & $38,000$ & 1.1 & $200$ \\
\CORE-150 & $52,000$ & 1.5 & $500$ \\
\CORE-180 & $65,000$ & 1.85 & $800$ \\
CMB-S4 (Atacama) & $10,700$ & 0.47 & $70$\\
CMB-S4 (South Pole) & $71,000$ & 6.9 & $5,000$\\
\CORE-150+CMB-S4 (Atacama) & $56,000$ & 2.5 & $850$\\
\CORE-150+CMB-S4 (South Pole) & $222,000$ & 21.5 & $20,000$\\
\hline
\end{tabular}
\caption{\label{tab:counts} Number of clusters $N_{\rm clus}$ expected to be detected with $S/N \geqslant 5$ for the experiments. For each experiment, the corresponding sky mask is taken into account. \CORE-150+CMB-S4 (Atacama)/(South Pole) counts are given within the Atacama/South Pole mask. For Atacama, the table gives the number of detections in the whole sky region covered by the Atacama survey. 
In the sky region not covered by the South Pole survey, i.e. 38\% of sky, the sole Atacama survey detects instead 7,400 clusters, 50 of which at $z>1.5$. The added value of a combination of \CORE\ with a deep, high-resolution ground-based survey is spectacular. }
\end{table}

\section{Science with the \CORE\ Cluster Sample}
\label{sec:analyses}

In the following subsections we present forecasts for particular scientific analyses that the \CORE\ dataset will enable.

\subsection{Cosmological Constraints from \CORE~Cluster Counts}
\label{sec:clustercounts}

This section presents cosmological constraints assuming the selection function based on thermal SZE noise maps computed in Section~\ref{sec:clustercats}.
We adopt a Markov Chain Monte Carlo (MCMC) approach, and we test our contours against those derived using a Fisher matrix technique in the case of constraints on ($\Omega_{\rm m}$, $\sigma_8$).

\begin{figure}[tbp]
\centering
\hbox to \textwidth{\hskip-1.5in
\vbox to 2.3in{\vskip-0.15in\includegraphics[width=0.55\textwidth]{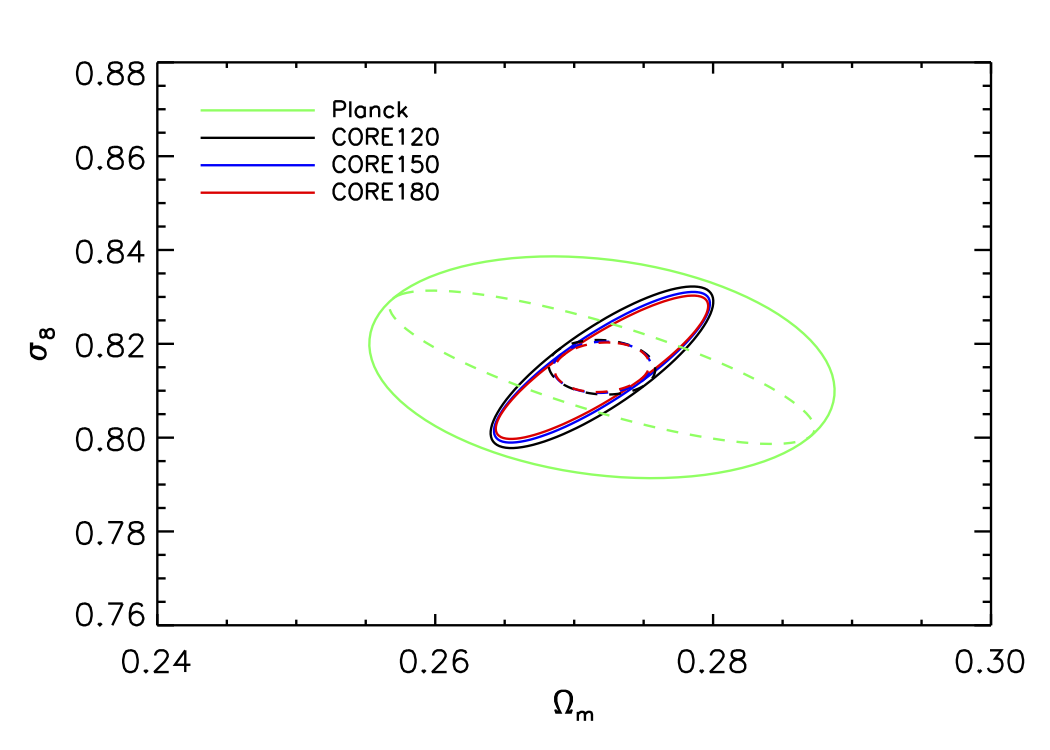} \vfil}
\hskip-3in\vbox to 2.5in{
\includegraphics[width=.45\textwidth]{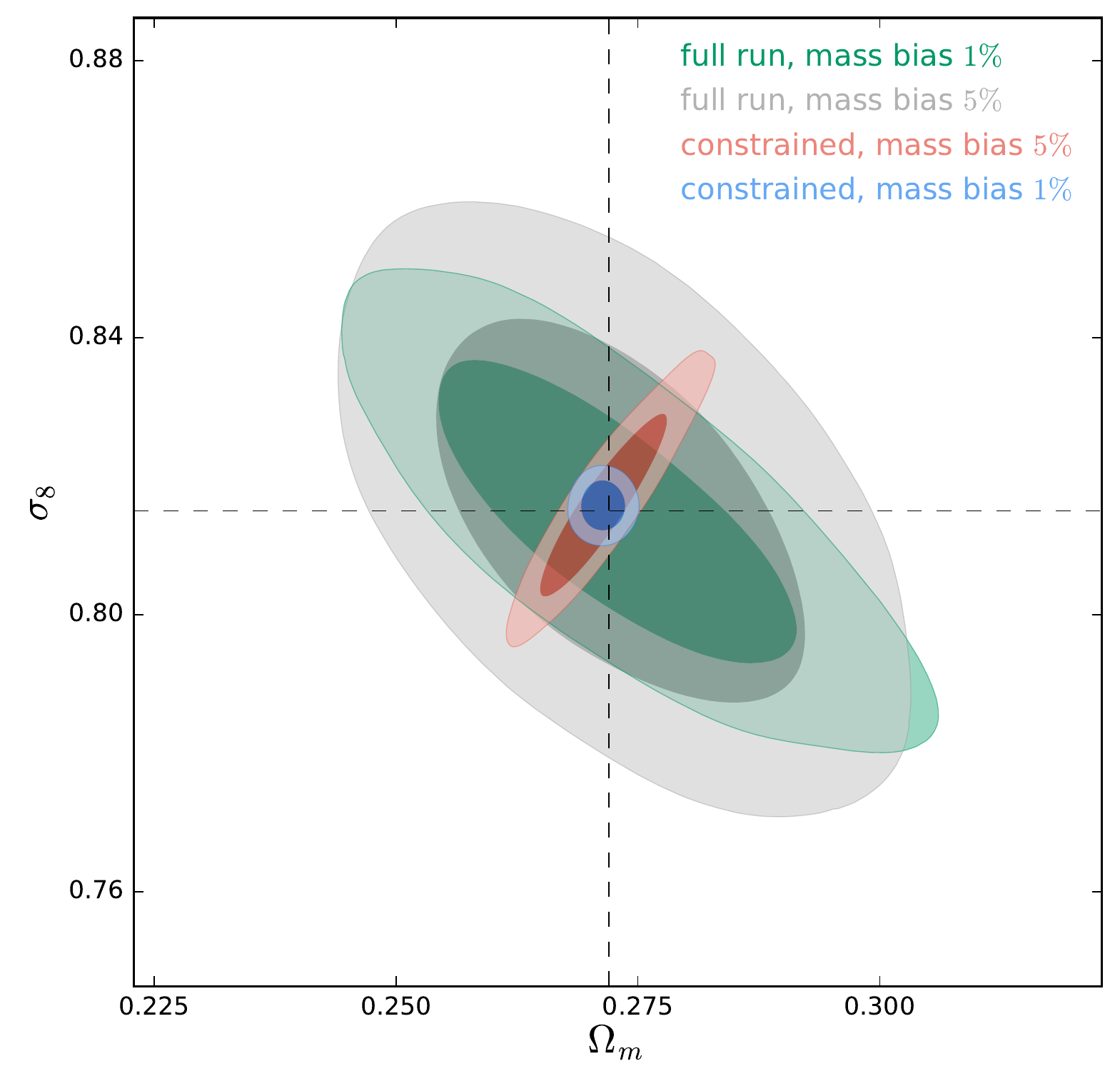}\vfil
\hfil}}
\caption{\label{fig:const1} {\it Left:} the 95\% confidence limits (from Fisher matrix analysis) on $\Omega_{\rm m}$ and $\sigma_8$ from \Planck\ and the three \CORE\ telescope apertures which we considered. The uncertainty on $1-b$ for each set of contours is 5\%/1\% for solid/dashed lines, respectively. {\it Right:} The 68\% and 95\% confidence limits (from MCMC analysis) for \CORE-150 in a constrained case where only $\Omega_{\rm m}$, $\sigma_8$ and $1-b$ are free to vary as in the Fisher analysis, and for the full run case where all the cosmological parameters and four mass--observable parameters are free and priors are adopted as in the \Planck\ analysis \cite{planck2014-a30}. The MCMC constrained case is in very good agreement with the Fisher blue solid and dashed contours of the left hand figure.}
\end{figure}

\subsubsection{Impact of Mass Calibration and Parameter Degeneracies}
\label{sec:sig8-omega}

We first focus on $\Omega_{\rm m}$ and $\sigma_8$. Although these two parameters will likely be constrained to a few parts in a thousand in the early 2020's after the end of dedicated large scale structure missions such as DESI and \Euclid~(see for example~\cite{sartoris2016}), we want to compare the gain in sensitivity from \Planck\ to \CORE\ using only SZE cluster counts. For this first test, we fix all the cosmological and scaling law parameters except $\sigma_8$, $\Omega_{\rm m}$ and $1-b$, where $b$ is a single fractional mass bias parameter assumed to hold for all masses and redshifts~\citep{planck2013xx}.  Although this approach does not allow for as much freedom as the cosmological analyses undertaken with the current datasets \citep{bocquet15,dehaan16}, the precise mass constraints from CMB lensing (see Section~\ref{sec:clustermass} and Figure~\ref{fig:cmblens_all} below) will dramatically reduce the systematic uncertainties and make it less important to include this additional freedom.

Results are shown in Figure~\ref{fig:const1} (left) for the Fisher matrix analysis of the cluster samples arising from the three different \CORE\ telescope sizes and from the \Planck\ sample in the case of 5\% (dashed line) and 1\% (solid line) priors on $1-b$.  For \Planck, improving our prior on $1-b$ from 5\% to 1\% reduces the contour perpendicularly to the well known parameter degeneracy as described in the recent \Planck\ analysis~\cite{planck2014-a30}. For \CORE, the orientation of the degeneracy line is similar to that for \Planck\ in the case of a 1\% calibration error but is different for a 5\% calibration error.  Note that the expected \CORE\ constraints are much stronger than the \Planck\ constraints (gain of factor 4 on $\Omega_{\rm m}$ and factor 3 on $\sigma_8$ for a 1\% prior on $1-b$). Interestingly, the three \CORE\ telescope sizes lead to comparable constraints when the error on $1-b$ is fixed through an external prior.  The significant difference between the 5\% and 1\% cases is indicating that the uncertainties on $\sigma_8$ and $\Omega_{\rm m}$ are dominated by the uncertainty on the cluster mass scale for \CORE, if that uncertainty is bigger than 1\%. This demonstrates the importance of having a mass calibration good to 1\% to be able to use the full information in the \CORE\ cluster counts.

The results from a MCMC analysis are shown in the right panel of Figure~\ref{fig:const1}.  As for the Fisher matrix case, we show results for the two different assumed accuracies of the mass calibration.  In addition, we show results for a constrained case (hereafter c-case) where only $\Omega_{\rm m}$, $\sigma_8$ and $1-b$ are free to vary as in the Fisher case, and for a full run case (hereafter f-case) where all the cosmological parameters are free and priors are adopted on the mass-observable relation parameters ($\log{Y_*}$, $\alpha$, $\beta$ and $\sigma_{\ln{Y}}$) as in the recent \Planck\ analysis \cite{planck2014-a30}.  The c-case is in good agreement with the Fisher matrix results on the left, whereas in the f-case the constraints are significantly broadened by the additional cosmological and mass-observable relation parameters.  In this f-case there is a clear advantage to having 1\% mass calibration, but the impact is less dramatic than for the c-case run.

\subsubsection{Impact of Cosmic Variance}

For this analysis, we have binned our sample in redshift assuming that cluster counts are uncorrelated, and we have employed a Poisson likelihood based on the Cash statistic~\cite{cash1979}. This approximation is valid for large redshift bins ($\Delta z \sim 0.1$) and medium size cluster samples such as \Planck's and SPT's where the statistical error budget is dominated by shot noise. With larger samples such as that from \CORE, we expect this approximation to break down due to large scale correlations in the underlying matter density field. The cluster counts within each bin $\Delta z_i$ then fluctuate due to the large scale structure~\citep{hucohn2006},
\begin{equation}
N_i = \bar N_i ( 1 + b_i \delta_i) \: ,
\end{equation}
where $\delta_i$ is the overdensity within the redshift bin, $b_i$ is the average cluster bias and we denote averaged quantities with overbars. If the bin size is large enough, the density fluctuations are Gaussian and fully characterised by their variance
\begin{equation}
\sigma^2(z_i) = \int \frac{\mathrm{d}^3 \mathbf k}{(2 \pi)^3} \: W^2(\mathbf k,z_i) P(k)
\end{equation}
with the window function $W(\mathbf{k},z_i)$ picking out radial shells around the observer. Because \CORE\ will observe a large number of clusters in each redshift bin, $\bar N_i \gg 1$, the likelihood to find $N^\mathrm{obs}_i$ objects is then given by a Gaussian with variance $s_i^2 = \bar N_i + \bar N_i^2 b_i^2 \sigma^2(z_i)$, receiving contributions from both shot noise and sample variance due to fluctuations in the density field~\citep{takadaspergel2014}:
\begin{equation}
\mathcal{L}_i = \frac{1}{\sqrt{2 \pi s_i^2}} \exp \left( \frac{(N_i^\mathrm{obs} - \bar N_i)^2}{2 s_i^2} \right) \: .
\end{equation}

\begin{figure}[tbp]
\centering
\includegraphics[width=.45\textwidth]{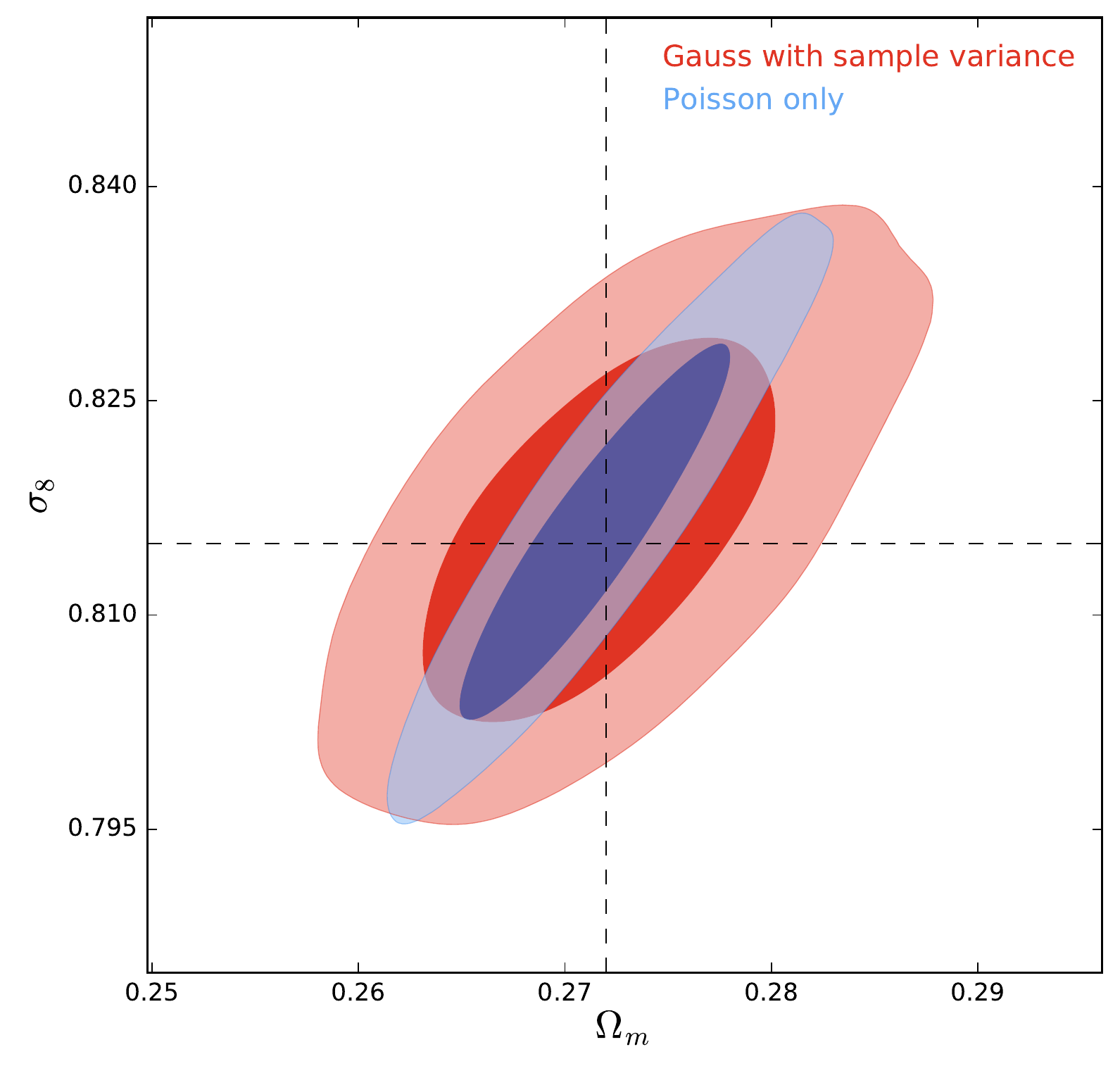} \includegraphics[width=.45\textwidth]{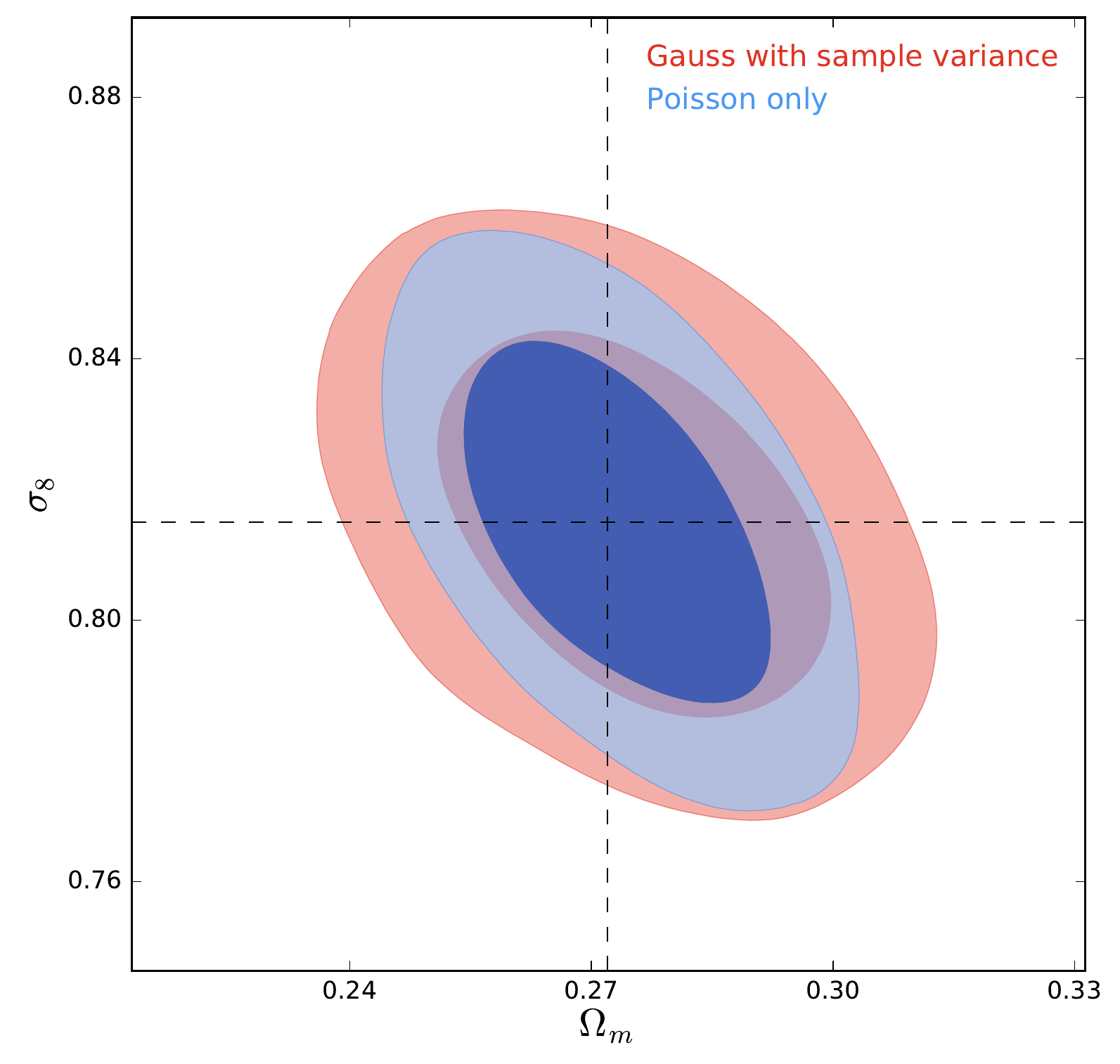}\\
\caption{\label{fig:const4} Impact of cosmic variance.  {\it Left: }In the c-case (see description in Figure~\ref{fig:const1} or Subsection~\ref{sec:sig8-omega}), the 68\% and 95\% confidence regions widen perpendicularly to the characteristic $\sigma_8$-$\Omega_{\rm m}$ degeneracy. {\it Right:} In the f-case, the impact is much smaller, because the constraints are already broadened due to the marginalization over the additional cosmological and mass--observable relation parameters. The two plots assume a 5\% calibration on the mass bias.}
\end{figure}

In Figure~\ref{fig:const4}, we compare the Poisson likelihood result to the Gaussian which takes into account the cosmic variance contribution. For the c-case (left), it widens the contour perpendicular to the usual $\sigma_8$-$\Omega_{\rm m}$ degeneracy direction, while for the f-case (right) the effect is smaller because statistical errors are less dominant when taking into account marginalization over additional cosmological and mass-observable relation parameters.

\subsubsection{Dark Energy Equation of State}

\begin{figure}[tbp]
\centering 
\includegraphics[width=.45\textwidth]{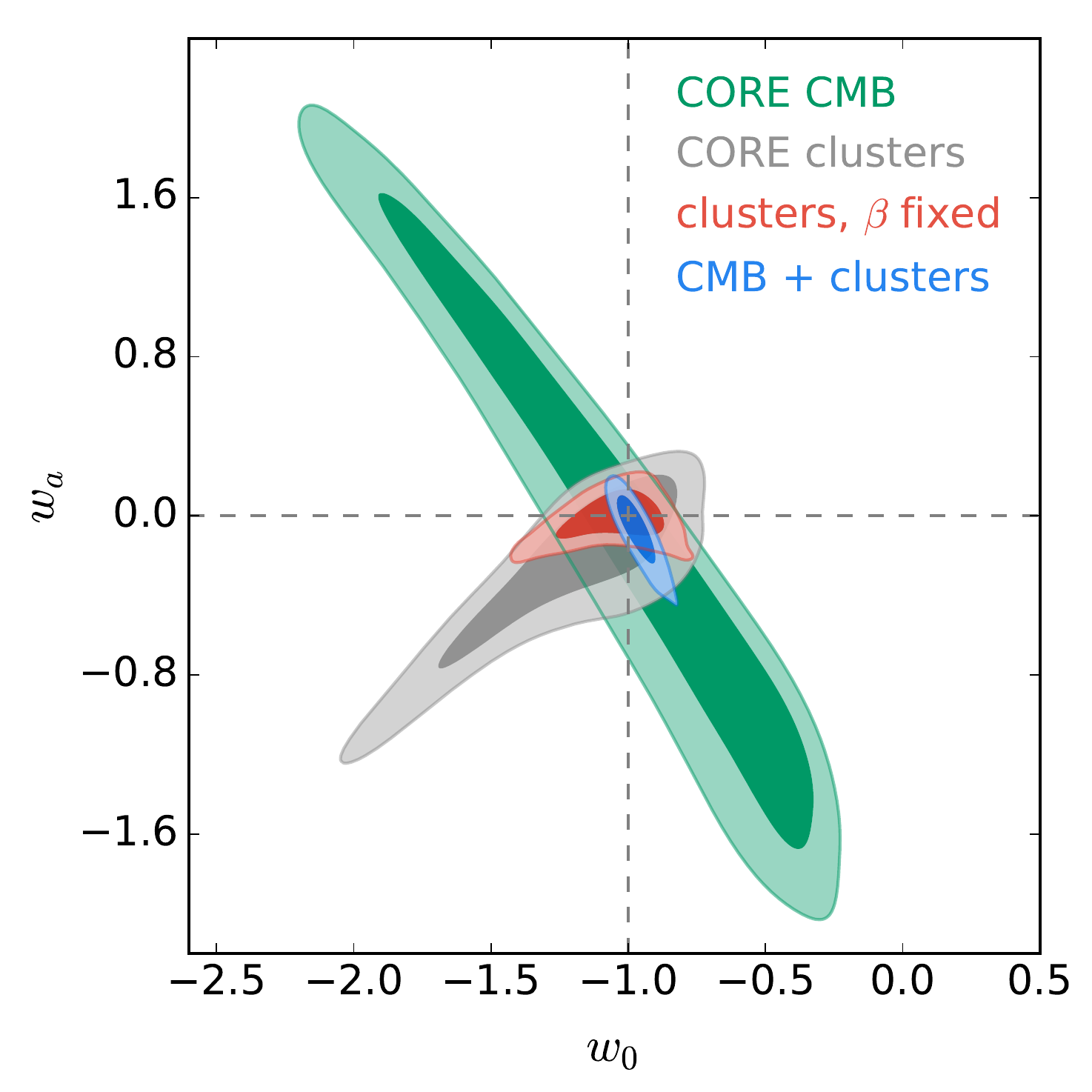}\\
\caption{\label{fig:const2} 68\% and 95\% confidence regions on $w_0$ and $w_a$ from \CORE-150 for cluster counts and primary CMB. Cluster counts break the degeneracy of the primary CMB constraints, providing joint constraints $\sigma_{w_0}=0.05$ and $\sigma_{w_a}=0.13$.}
\end{figure}

We now study the $w$CDM model with dark energy equation of state parameters $(w_0, w_a)$ using $w(a)=w_0+w_a \times (1-a)$. We leave all the cosmological parameters free and adopt the same priors on the mass observable relation parameters ($\log Y_*$, $\alpha$, $\beta$, $\sigma_{\ln Y}$) as in the recent analysis of the \Planck\ sample~\cite{planck2014-a30}. For this specific parameter combination, knowing the cluster mass scale parameter $b$ is less important than for the previous case ($\Omega_{\rm m}$ vs. $\sigma_8$), because $w_0$ and $w_a$ constraints are mainly dependent on the evolution of the cluster counts and are less sensitive to their overall normalization. Constraints on $(w_0,w_a)$ are shown in Figure~\ref{fig:const2}. Using cluster counts only, we 
forecast fully marginalized 68\% confidence constraints of $\sigma_{w_0}=0.28$ and $\sigma_{w_a}=0.31$ for our \CORE-150 baseline configuration. Constraints from \CORE\ primary CMB only are limited to $\sigma_{w_0}=0.47$ and $\sigma_{w_a}=0.97$. Combining the \CORE\ primary CMB constraints to \CORE\ cluster counts breaks the degeneracy and provides $\sigma_{w_0}=0.05$ and $\sigma_{w_a}=0.13$. If the redshift trend parameter $\beta$ is known, cluster counts  constraints tighten to $\sigma_{w_0}=0.13$ and $\sigma_{w_a}=0.10$. These constraints are competitive and complementary to the constraints expected from weak lensing and galaxy clustering in the 2020's: \Euclid\ forecasts $\sigma_{w_0}=0.015$ and $\sigma_{w_a}=0.15$~\citep[Table 2.2 of][]{laureijs11} and LSST $\sigma_{w_0} \sim 0.05$ and $\sigma_{w_a} \sim 0.15$~\citep[Fig. 15.1 of][]{LSST09}.\\

\begin{figure}[tbp]
\centering
\includegraphics[width=.35\textwidth]{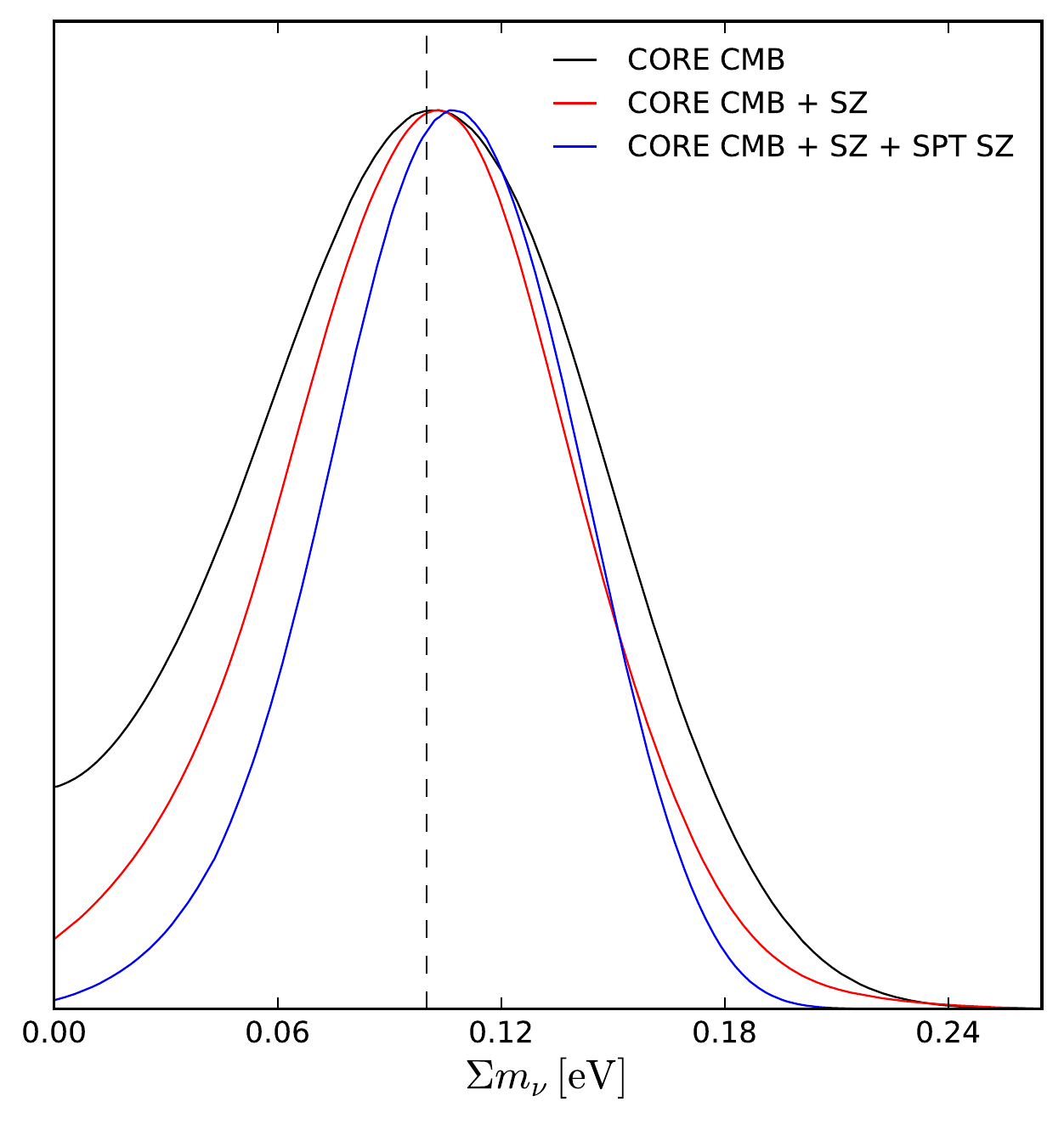} \includegraphics[width=.6\textwidth]{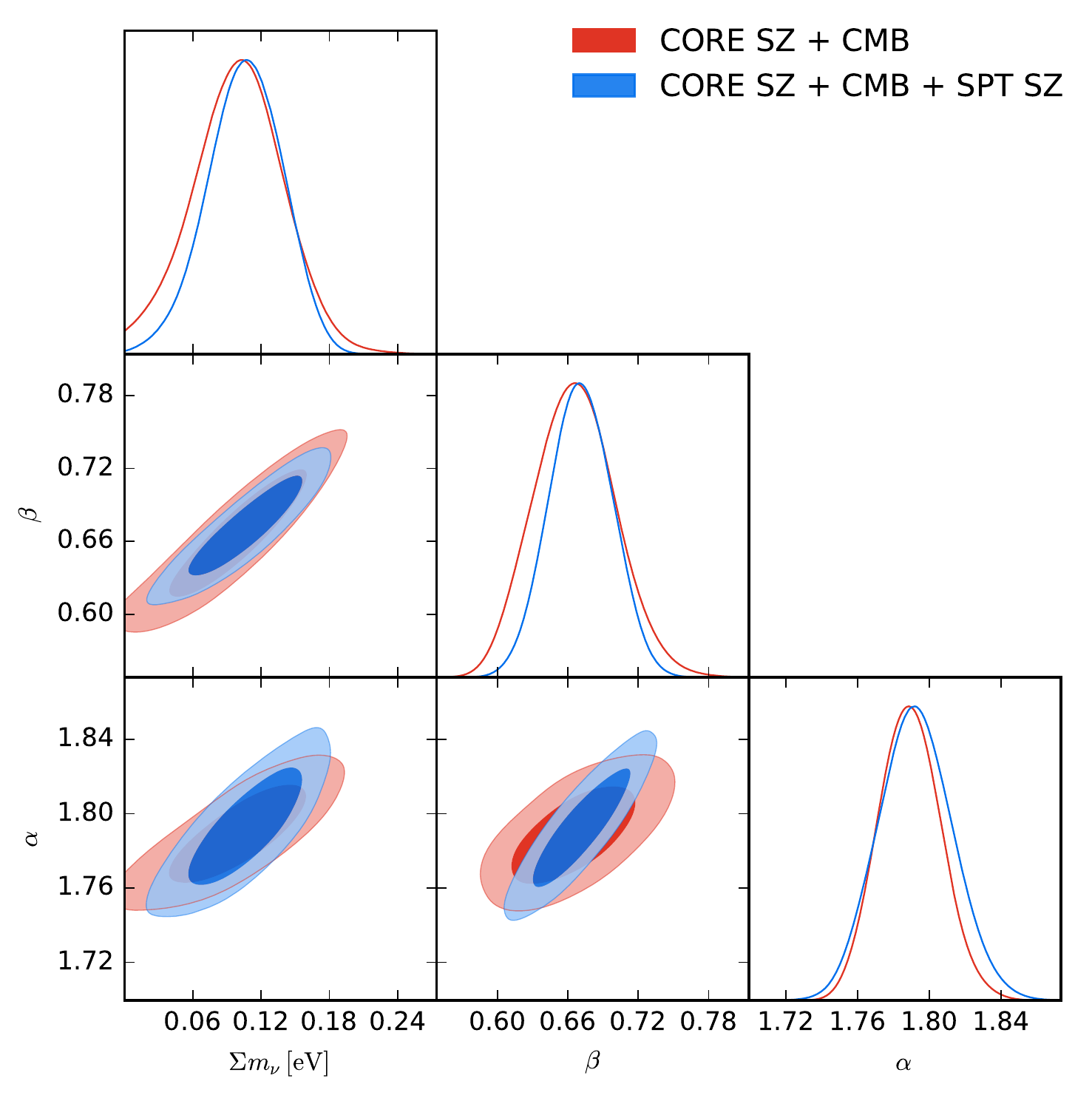}\\
\caption{\label{fig:const3} {\it Left:} Probability distribution function of $\Sigma m_\nu$ for \CORE-150 primary CMB (TT,TE,EE), \CORE-150 CMB+SZ, and \CORE-150 CMB+SZ combined with South Pole SZ. {\it Right:} Degeneracies between $\Sigma m_\nu$ and the slope and evolution parameters ($\alpha$, $\beta$) of the tSZE flux-mass relation.}
\end{figure}

\subsubsection{Neutrino Mass Constraints}

tSZE cluster counts alone cannot provide competitive constraints on the sum of the neutrino masses $\Sigma m_\nu$, because the mass sum is degenerate with the normalization of the primordial power spectrum. Combining \CORE\ primary CMB and \CORE\ cluster counts strengthens significantly the constraints on this parameter in the $\Lambda$CDM+$\Sigma m_\nu$ model.  To explore this, we use the chains from the Exploring Cosmic Origins paper on cosmological parameters~\cite{coreparam2016} in combination with our cluster MCMC. Figure~\ref{fig:const3} (left) presents the probability distribution function of $\Sigma m_\nu$ for \CORE-150  primary CMB TT, TE and EE (solid black line), \CORE-150 primary CMB + cluster counts (solid red line) and  \CORE-150 primary CMB + cluster counts in combination with CMB-S4 (South Pole) cluster counts (solid blue line). We obtain the following constraints on the sum of the neutrino masses $\sigma_{\Sigma m_\nu}=47$, 39, and 33~meV for \CORE-150 CMB, \CORE-150 CMB+SZ and \CORE-150 CMB+SZ + CMB-S4 (South Pole), respectively. Figure~\ref{fig:const3} (right) presents the degeneracies of $\Sigma m_\nu$ with the mass-observable scaling relation parameters $\alpha$ and $\beta$. Improving our knowledge of cluster masses would strengthen further our constraints on the sum of the neutrino mass.
 
\subsubsection{Information Gain}\label{sec:info_gain}

\begin{figure*}

\includegraphics[width=\textwidth]{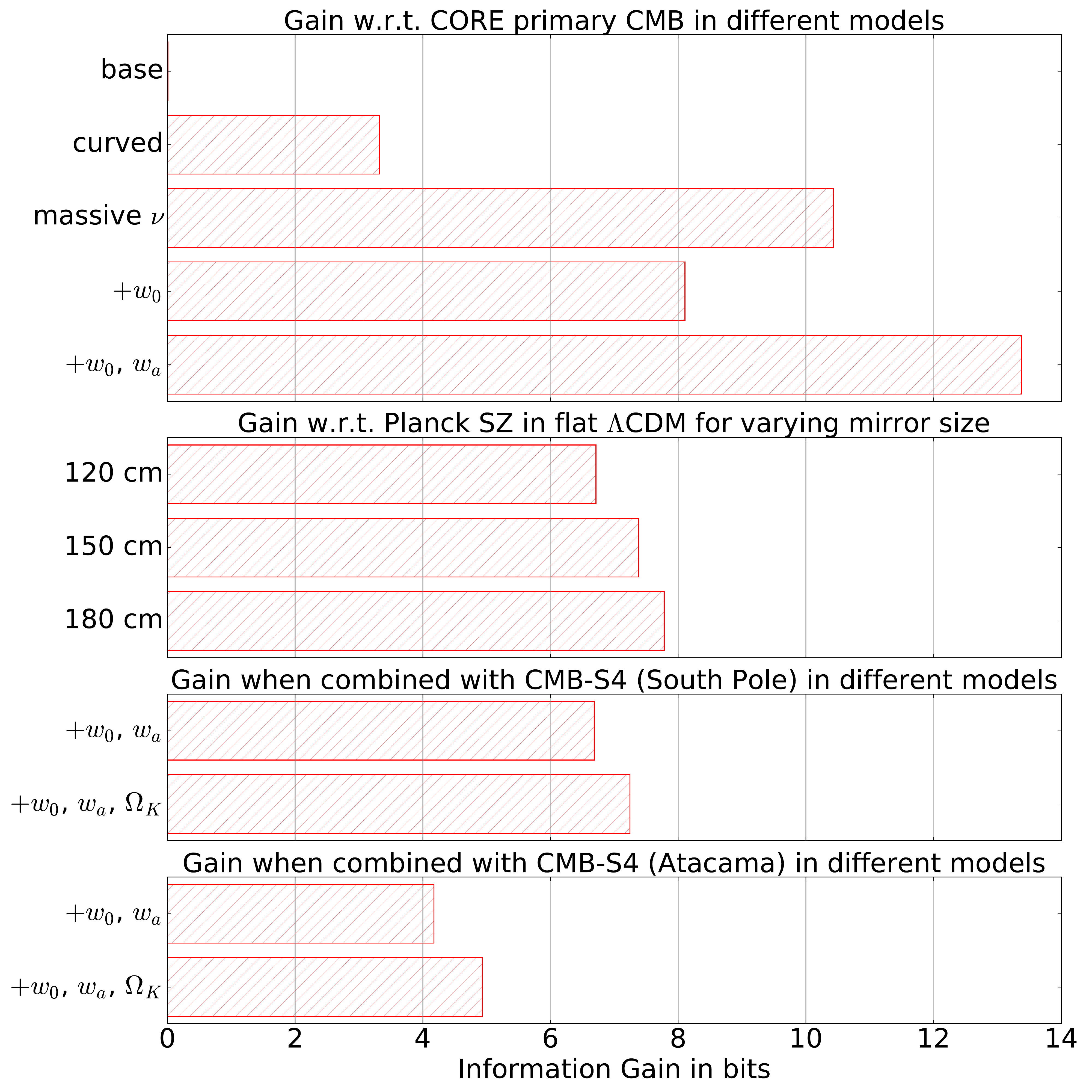}
\caption{Information Gain for the \CORE\ SZE number counts w.r.t. different priors in different models. From top to bottom, we find a strong synergy between \CORE\ primary CMB and \CORE\ SZE in models with massive neutrinos and dynamical Dark Energy. Furthermore, the information gain of \CORE\ SZE relative to \Planck\ SZE is comparable to the 7.6 bits of information gained by moving from WMAP 9 primary CMB to \Planck\ 2015 primary CMB \cite{SG}. Finally, \CORE\ SZE will provide useful extra information when combined with ground based CMB-S4 experiments.}

\label{fig:info_gains}
\end{figure*}
%

To evaluate the performance of a future experiment and how it is affected by specific choices such as the mirror size, Figures of Merit (FoM) are employed.  Recently, in addition to the traditional Dark Energy Task Force (DETF) FoM \cite{DEFT} and variations thereof, the so--called \emph{Information Gain} has been introduced and applied to cosmology \cite[see e.g.][]{bayes_in_cosmo, march, modelbreaking, seehars1, Paykari, SG, info_gain_raveri}.  Given a prior covariance $\Pi$, and a posterior covariance $\Sigma$, the Information Gain can be computed as \cite{bayes_in_cosmo, Paykari}: 

\begin{equation}\label{eq:info_gain}
\mathcal{I} = \frac{1}{2} \ln\Big(\frac{\text{det}(\Pi)}{\text{det}(\Sigma)}\Big) -\frac{1}{2} \text{trace} (\mathbb{I} - \Sigma\Pi^{-1}),
\end{equation}
where $\mathbb{I}$ is the identity matrix. 

The unit of the information gain depends on the base of the logarithm used in its derivation.  If the natural logarithm is used, as in Eq.~\ref{eq:info_gain}, it is `nats'; if Eq.~\ref{eq:info_gain} is divided by $\ln(2)$, the unit is the more familiar `bits'\footnote{
Consider a single free parameter in a Poisson system.  In this case $\Sigma=\Pi/\lambda$ where $\lambda$ is the ratio of the number of samples in the prior and posterior case.  For an increase in sample size of an order of magnitude, the number of bits of information would be 8.2.},
which corresponds to using the logarithm base 2, which is what we use in this work.

Contrary to the traditional DETF FoM, which considers the determinant of the Fisher matrix, the information gain is motivated by information theory \cite{kullbackleibler, info_theory}, and quantifies the amount of information on the model parameters that is gained when updating the prior to the posterior (for application to past observations, see \cite{seehars1, SG}). For this reason, the information gain has been proposed as a FoM for experimental forecasting \cite{bayes_in_cosmo, modelbreaking, Paykari}. The advantages of the information gain compared to other FoM are discussed in detail elsewhere \cite{bayes_in_cosmo, Paykari}.

Let us first consider the information gained when \CORE\ SZE cluster counts are combined with the cosmological constraints from the primary \CORE\ CMB in different cosmological models. We consider the base flat $\Lambda$CDM model, the curved $\Lambda$CDM, the flat $\Lambda$CDM with massive neutrinos, and the flat $\Lambda$CDM with dynamical Dark Energy Equation of state. In each of these models, we use the primary CMB Fisher matrices~\cite{coreparam2016} as priors, and investigate how things improve when one includes the SZE number counts. The results are shown in the uppermost panel of Figure~\ref{fig:info_gains}.

In flat $\Lambda$CDM, \CORE\ SZE provides very little information when added to primary \CORE\ CMB, as the latter already constrains the cosmological parameters to better than percent accuracy. Therefore, in this model, assessing the consistency between \CORE\ SZE and \CORE\ primary CMB will be a valuable cross check. This situation changes when extended models are considered. When curvature is allowed, the addition of \CORE\ SZE gives a modest improvement of 3.3 bits, mainly due to the improvement of the constraint on the curvature density. However, much larger Information Gains are expected in models with massive neutrinos (7.5 bits), and even more so in models with dynamical Dark Energy. In this model, the addition of \CORE\ SZE will provide a boost of 13.4 bits of information. This results from the ability of \CORE\ SZE constraints to break the well known parameter degeneracies of the primary CMB present in these models. In summary, the synergy between \CORE\ cluster counts and primary CMB manifests itself most in the ability to dramatically improve constraints on neutrino masses and the Dark Energy Equation of state parameters.

We also investigate the information gain of moving from the \Planck\ 2015 SZE number counts to the \CORE\ SZE number counts, as a function of the mirror size. For this purpose, in Eq.~\ref{eq:info_gain} we assume the \Planck\ SZE covariance as a prior $\Pi$, and the \CORE\ SZE Fisher matrix as a posterior $\Sigma$. The results are shown in the second panel of Figure~\ref{fig:info_gains}. Moving from \Planck\ SZE number counts to \CORE\ SZE number counts will provide an information gain of 6.7-7.8 bits, depending on the mirror size. This corresponds to a large amount of additional cosmological information, comparable to the improvement obtained in moving from the WMAP 9 primary CMB results to the \Planck\ 2015 primary CMB results \cite{SG}. Naturally, the mirror size affects the information gain, with a larger mirror resulting in more detected objects and therefore more cosmological information. This effect is, however, not linear in mirror size: reducing the mirror size by 30 cm (from 150 cm to 120 cm), results in a loss of $\sim$ 0.7 bits of information, whereas increasing the mirror size by 30 cm (from 150 cm to 180 cm), yields a smaller difference of 0.4 bits. 

Besides the noticeable improvement \CORE\ SZE would provide relative to \Planck\ SZE, and its synergies with \CORE\ primary CMB, we also investigate how much information \CORE\ SZE would provide when added to the ground-based SZE experiments CMB-S4 (South Pole) and CMB-S4 (Atacama). In this case, we assume the CMB-S4 Fisher matrices as priors, and the combination of these with \CORE\ SZE as posterior. We fix the mirror size to 150 cm. The results are summarized in the two lower panels of Figure~\ref{fig:info_gains}. Considering the large information gains obtained by adding \CORE\ SZE to the ground based experiments (6.7-7.3 bits with CMB-S4 (South Pole), 4.1-4.9 bits with CMB-S4 (Atacama)), we conclude that \CORE\ would significantly improve the cosmological constraints of ground based experiments and provide a good amount of new cosmological information on models of dynamical Dark Energy with and without free curvature. Furthermore, we find a larger information gain when \CORE\ SZE is added to CMB-S4 (South Pole) rather than CMB-S4 (Atacama). This effect is due to the fact that the joint CMB-S4 (South Pole)+\CORE\ sample is in itself larger, and, furthermore, spans a larger portion of mass--redshift space (see Figure \ref{fig:mz} and Table \ref{tab:counts}).

\subsection{Cluster Mass Calibration}
\label{sec:clustermass}

Cluster masses will be calibrated using lensing of the primary CMB by clusters, a technique called CMB Halo Lensing. \Planck, ACT and SPT recently reported first detections of this effect by stacking hundreds to thousands of objects~\cite{planck2014-a30,madhavacheril2015,baxter2015}.  In this section we provide forecasts for the three \CORE\ concepts we consider (\CORE-120, \CORE-150, \CORE-180) based on a new detection method \citep{melin2015}.

\begin{figure}[tbp]
\centering
\includegraphics[width=.45\textwidth]{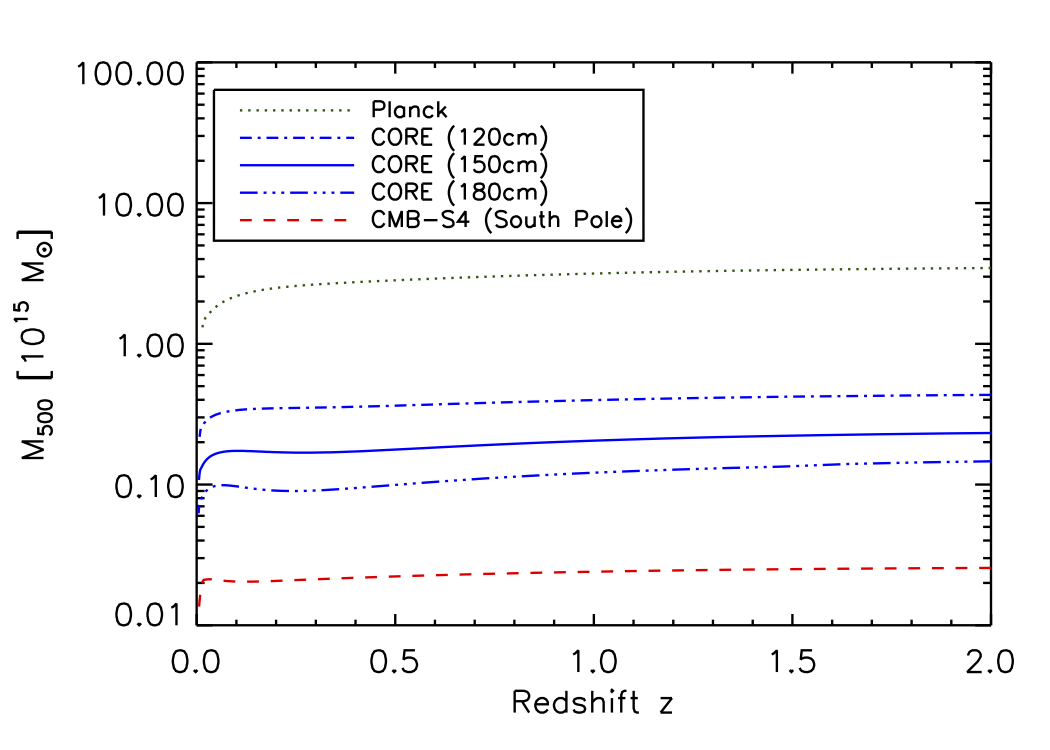} \includegraphics[width=.45\textwidth]{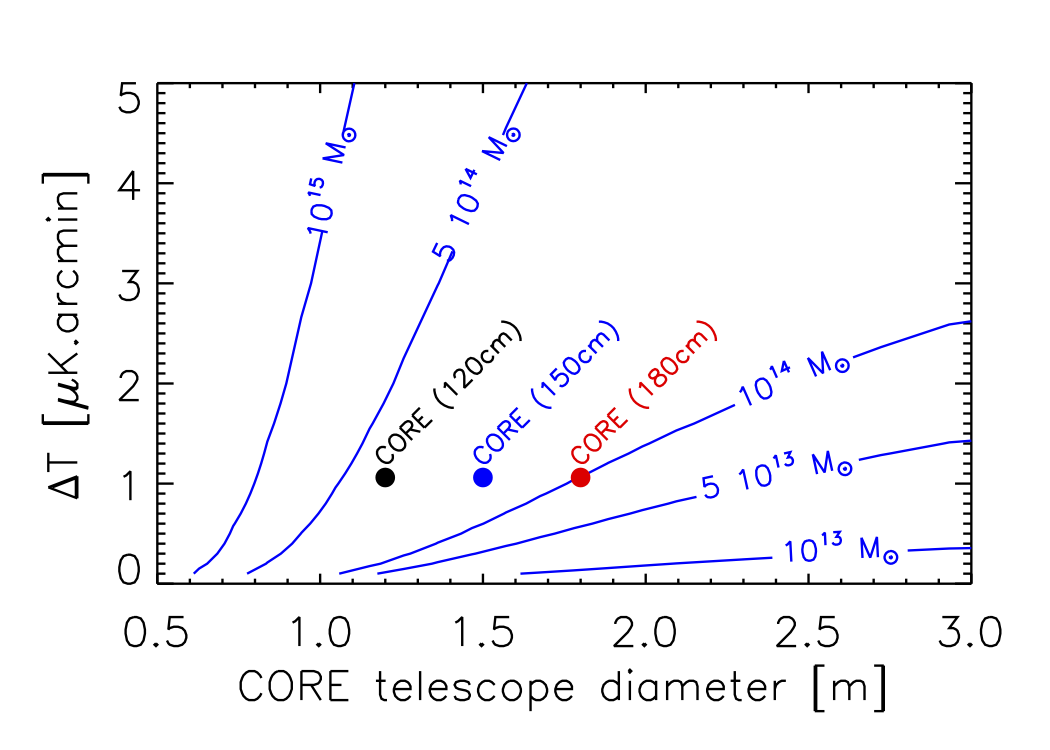}\\
\caption{\label{fig:cmblens} {\it Left:} $1\sigma$ uncertainties on CMB halo lensing mass as a function of redshift for \Planck, for the three \CORE\ configurations and CMB-S4 (South Pole). {\it Right:} $1\sigma$ uncertainties on the mass as a function of instrumental noise and telescope aperture (at $z=0.5$). High resolution (180~cm aperture) is required to achieve single halo mass uncertainties of $10^{14} \, M_\odot$ via halo CMB lensing.}
\end{figure}

Figure~\ref{fig:cmblens} left shows the $1\sigma$ error on CMB halo lensing mass as a function of redshift for \Planck\ (green dotted line), the three \CORE\ concepts (blue lines) and the CMB-S4 (South Pole; red dashed line) experiment. While detecting individual cluster mass was out of reach for \Planck, \CORE-120/\CORE-150/\CORE-180 will be able to provide individual cluster masses with 1$\sigma$ statistical uncertainties of $4\times10^{14}/2\times10^{14}/ 10^{14} \, M_\odot$, respectively.
CMB-S4 (South Pole) could reach the $1-2 \times10^{13} \, M_\odot$ mass range if the experiment has sufficient frequency coverage to separate the SZE from point sources in clusters.
The right panel of Figure~\ref{fig:cmblens} presents mass isocurves at $z=0.5$ ($1\sigma$) for varying \CORE\ telescope apertures and instrumental noise levels. It illustrates in particular how quickly we gain in sensitivity on cluster mass when increasing the telescope aperture.\\

\begin{figure}[tbp]
\centering
\includegraphics[width=.45\textwidth]{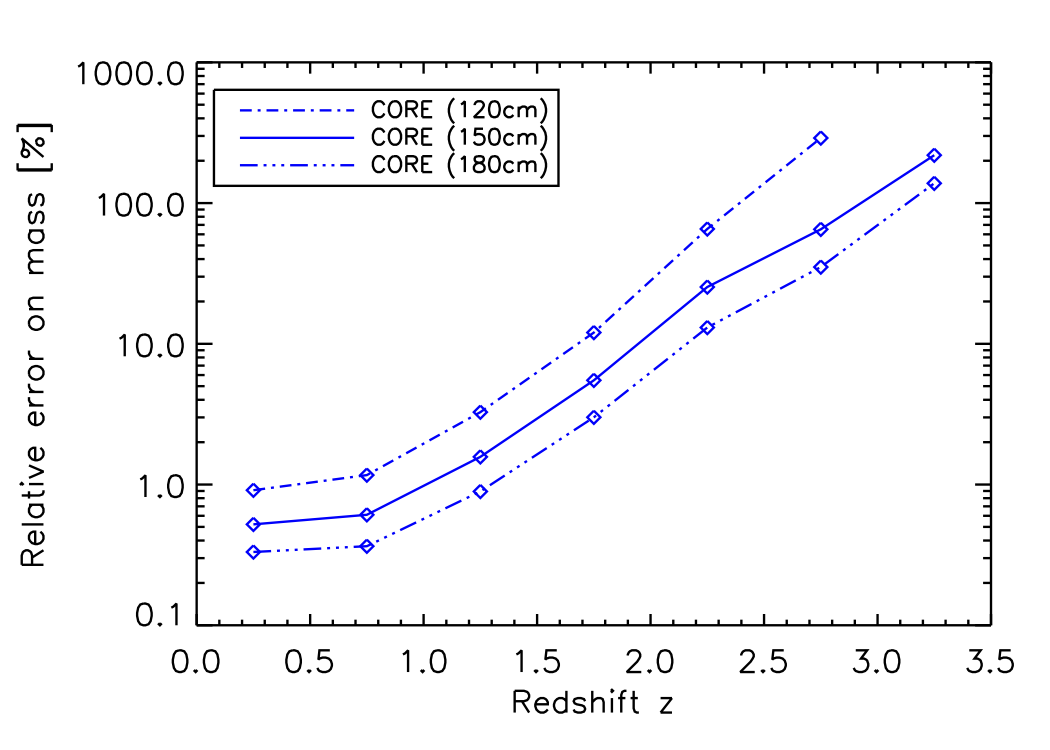} \includegraphics[width=.45\textwidth]{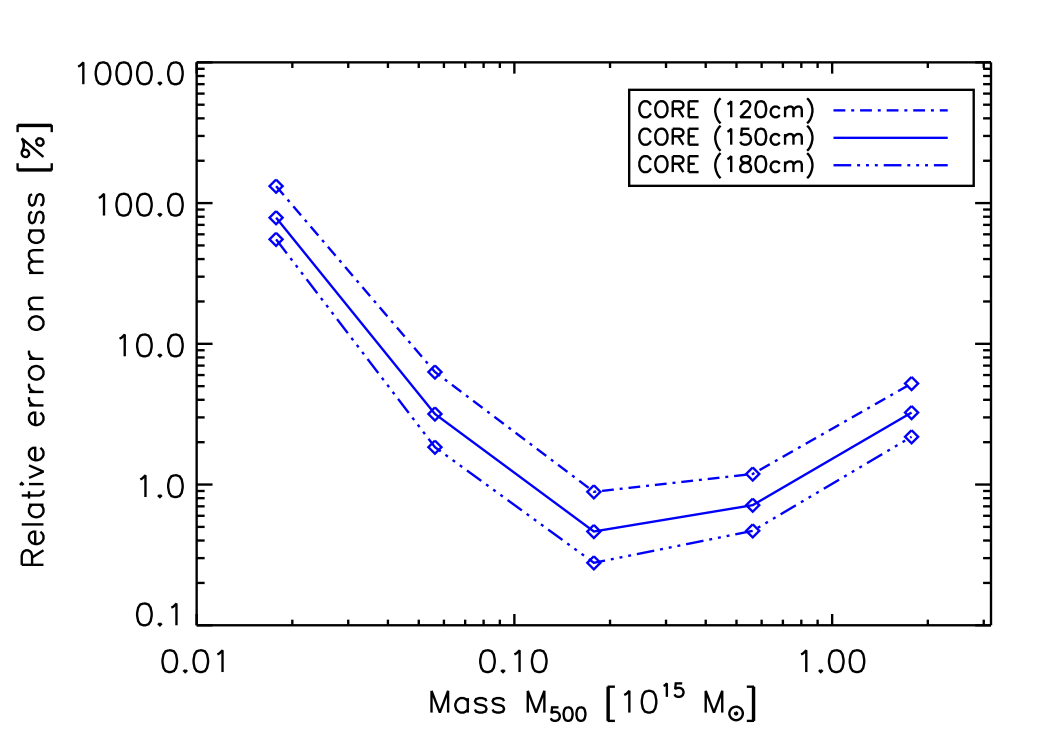}
\caption{\label{fig:cmblens_all} {\it Left:} Relative error on the mass scale $1-b$ from CMB halo lensing as a function of redshift for the three \CORE\ configurations. The same quantity from the stacked signal of the detected clusters in each redshift bin (Table~\ref{tab:counts}). {\it Right:} Relative error as a function of mass.}
\end{figure}

CMB halo lensing stacks will be used to calibrate cluster mass in scaling relations (e.g. $Y-M$) to allow for improved cosmological constraints.  Assuming that $1-b$ does not depend upon redshift or mass, \CORE-120/\CORE-150/\CORE-180 will constrain this parameter to 0.7\%/0.4\%/0.2\% (statistical error) if we stack all the clusters detected at $S/N \geqslant 5$ (numbers given in Table~\ref{tab:counts} for each concept). If $1-b$ depends on the redshift, \CORE-120 can still constrain it at the few percent level up to $z=1.5$, while \CORE-150 and \CORE-180 should be able to reach this precision up to $z=2$ (Figure~\ref{fig:cmblens_all} left). The right panel of Figure~\ref{fig:cmblens_all} shows that the mass dependence can also be constrained at the few percent level for clusters with masses between  2 and $3\times10^{13} \, M_\odot$ using stacking.\\

These numbers are competitive with-- and complementary to-- forecasts for \Euclid\ shear measurements (1\% calibration of the mass) in the $z<1$ redshift range. \CORE\ will provide access to lensing halo mass calibration at a few percent level for redshifts higher than $z=1$ (and up to $z=2$ for telescope sizes greater than 1.5\,m).

\subsection{Studies of the Relativistic Thermal Sunyaev-Zel'dovich effect}
\label{sec:relativistic}

We also explore the possibility of measuring the relativistic thermal SZE~\citep{itoh1998,sazonov1998,challinor1998} with \CORE-150.  This is particularly interesting to constrain cluster temperature without external X-ray follow-up. The analysis is based on the 100 highest temperature clusters in the \CORE\ sample. All these systems have temperatures at or above $k_B$T=12.6\,keV.  After extracting square patches of 2.5 degrees (on a side) around each cluster and for each band, the maps above 115\,GHz are smoothed to a common resolution of 7\,arcminutes. The maps at 115\,GHz and below are left at their native resolution. 

The maps are then preprocessed to remove point sources, Galactic emission and the primary CMB. Point sources are subtracted based on the difference map between the 60\,GHz and the 70\,GHz map (radio sources) and the 800\,GHz map (IR sources). A Mexican Hat Wavelet filter is applied to detect and mask IR sources while the radio sources are identified as peaks ($\ge4\sigma$)  in the difference map (60-70\,GHz). To remove the contribution from the thermal dust emission of the Galaxy, a modified black body with spectral index $\beta=1.6$ and variable temperature is fit to the dust spectrum. The spectrum of the dust is obtained as the mean flux beyond a region of 20\,arcminutes away from the cluster position (and up to the boundary of the field of view of 2.5\,degrees). A consistent best fit of 18\,K is found for all the cluster regions.
Only the bands above 390\,GHz (inclusive) are used to fit the spectral energy distribution of the thermal dust emission. The normalization of the model is taken such that the 800\,GHz band is reproduced exactly by the model. This model ($\beta=1.6$, $T=18$\,K, Norm=800\,GHz) is subtracted from all bands. Finally, the CMB is removed by subtracting the cleaned 220\,GHz band (where point sources are masked and the Galactic emission has been subtracted). The resulting SZE maps are dominated by SZE signal but potentially also contain some residual point source and IR emission (Galactic and extragalactic) that was not perfectly removed as well as instrumental noise.  An advantage of the \CORE\ frequency coverage is the ability to make this correction quite accurately.   

\begin{figure}[tbp]
\centering 
\includegraphics[width=.55\textwidth]{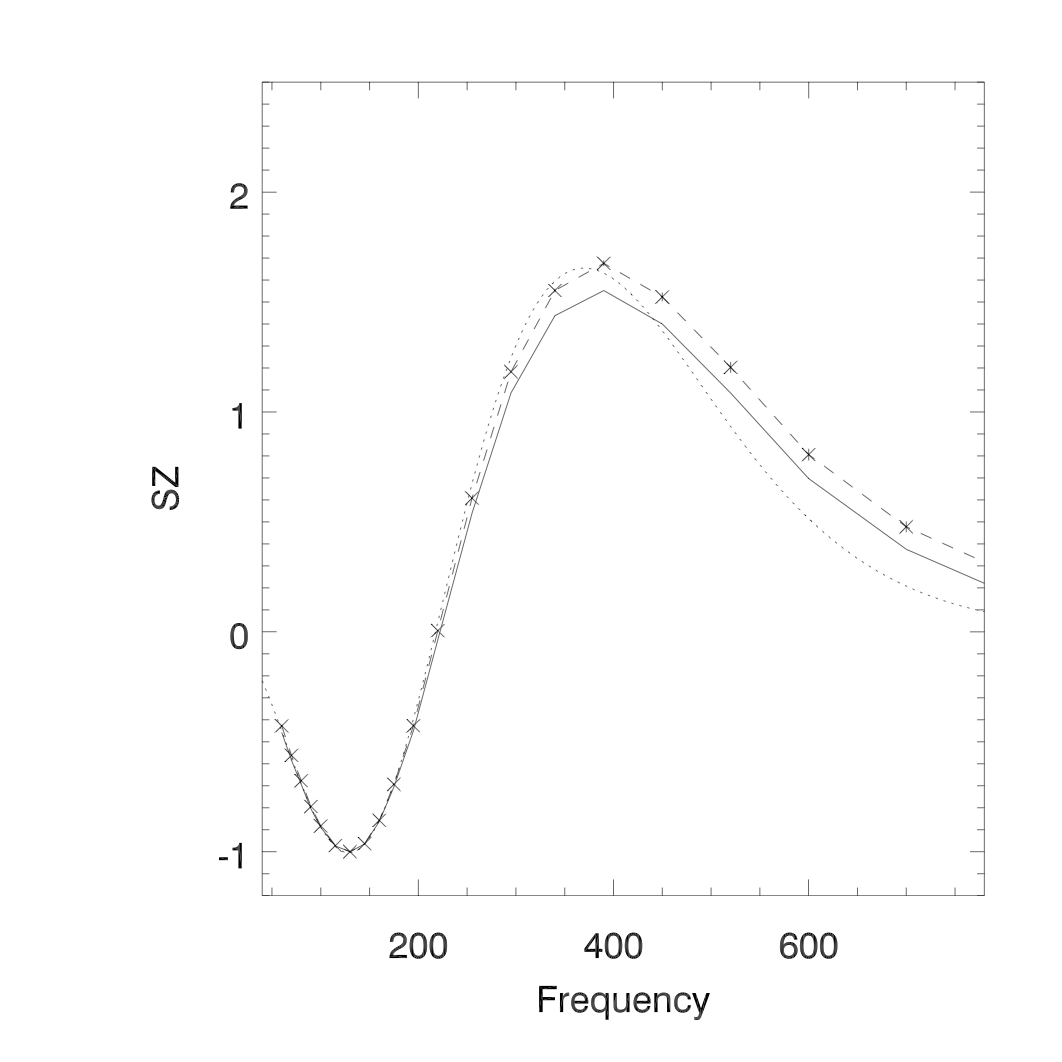}
\vskip-0.25in
\caption{\label{fig:relat}  Stacked, normalized SZE spectrum measurements (symbols) and uncertainties for the 100 highest temperature clusters. All spectra are normalized at 130\,GHz before stacking.  The lines shows the average of the stacked SZE spectrum (dashed),  an SZE spectrum  with no relativistic correction (dotted), and an SZE spectrum with relativistic correction (solid) for a cluster with $k_B$T=12\,keV, which is the minimum temperature for this stack.  The model spectra are also normalized at 130\,GHz.  The depth of the \CORE\ data enable the precise measurement of the cluster SZE spectrum even for much smaller numbers of clusters.}
\end{figure}

We use these maps to measure the SZE spectrum. The SZE flux is estimated in each band from the SZE maps as the minimum/maximum flux in a disc of 6~arcminutes centered on the cluster minus the average signal in a ring with inner radius equal to 30\,arcminutes and outer radius equal to 60\,arcminutes. The error on the flux is taken as the dispersion of the entire field of view after excluding the cluster region divided by the square root of the number of samples.  The measured SZE spectrum is finally used to derive the best temperature taking advantage of the dependency of the relativistic correction on the cluster temperature. Fitting the SZE spectrum is done excluding the bands below 130\,GHz (poor resolution and sensitivity), the band at 220\,GHz (used to subtract the CMB component) and the band at 800~GHz (used to remove the Galactic component). The approach is highlighted in Figure~\ref{fig:relat}.

The stacked normalized SZE spectrum for 100 stacked clusters (symbols) shows an excess at high frequencies with respect to the expected SZE spectrum (solid line) and the SZE spectrum without a relativistic correction (dotted line).  This excess is due to the relativistic model being for a 12.6\,keV cluster, which is the minimum temperature of the sample rather than the best fit mean temperature.  The measurement uncertainties are exceedingly small and reflect the uncertainty on the mean when considering the cluster to cluster spectral variation.  It is clear that even with a small number of clusters the \CORE\ data will enable precise measurements of the SZE spectrum.  Crucially important in this analysis is the ability to remove infrared source contamination, and that requires the broad frequency range of \CORE\ and especially the high frequency bands that are only available in a space mission.

\subsection{Constraints on the Kinetic Sunyaev-Zel'dovich Effect}
\label{sec:kSZE}

In this section, we present forecasts on the sensitivity of the \corep~mission for studies of the kinetic SZE \citep[][]{kSZ}. This effect expresses the Doppler kick experienced by CMB photons as they Thomson scatter off moving clouds of ionized electrons. This process results in a perturbation of the brightness temperature of the CMB that is proportional to the line-of-sight (LOS) component of the electron cloud peculiar velocity \citep[][]{kSZveryfirst}:
\begin{equation}
\frac{\delta T(\vn)}{T_0} = -\int dl\, \sigma_T n_e(\vn,l) \frac{\vv_e\cdot\vn}{c},
\label{eq:kSZ1}
\end{equation}
where $n_e$ is the electron number density, $\sigma_T$ is the Thomson cross-section, $\vv_e/c$ is the peculiar velocity vector of the cloud in units of the speed of light, and $\vn$ is the unit vector defining  the LOS. 

In our forecasts we study the \CORE\ sensitivity to the kSZE pairwise momentum $p_{kSZ}(r)$ on our set of galaxy clusters, which is given by this sum over galaxy cluster pairs:
\begin{equation}
\hat{p}_{\rm kSZ} (r) = -\frac{\sum_{i<j}(\delta T_{ i} - \delta T_{ j} )\,c_{ i,j}}{\sum_{ i<j} c_{ i,j}^2},
\label{eq:pksz1}
\end{equation}
 where $\delta T_i$ are the kSZE temperature estimates of the $i$-th cluster and the weights $c_{ i,j}$ are given by \cite{ferreiraetal99}
\begin{equation}
c_{ i,j} = \hat{\vrv}_{ i,j} \cdot \frac{\hat{\vrv}_{ i}+\hat{\vrv}_{ j}}{2} = 
 \frac{(r_{ i}-r_{ j})(1+\cos\theta)}{2\sqrt{r_{ i}^2 + r_{ j}^2 - 2r_{ i}r_{ j}\cos\theta}}.
\label{eq:cweight}
\end{equation}
As discussed in detail elsewhere \cite{Handetal2012}, ${\vrv}_{ i}$ and ${\vrv}_{ j}$ are the comoving distance vectors corresponding to the locations of the $i$-th and $j$-th clusters on the sky, and ${\vrv}_{ i,j} = \vrv_{ i}-\vrv_{ j}$ is the vector pointing to cluster $i$ from cluster $j$.  The symbol $\hat{\vrv}$ denotes a unit vector in the direction of $\vrv$, and $\theta$ is the angle separating the two directions $\hat{\vrv}_{ i}$ and $\hat{\vrv}_{ j}$. Similarly, $\hat{\vrv}_{i,j}$ is the unit vector defining the direction of the separation vector $\vrv_{i,j}$. 

For the theoretical modeling of this statistic, we follow an approach \cite{soergel16} where the pairwise momentum is defined as
\begin{equation}
p_{kSZ}(r) = T_0\, \tau_{eff}\, \frac{2\,b_{\rm cl} \xi^{\delta\, v}(r,z)}{1+b_{\rm cl}^2\xi^{\delta \delta}(r,z)}, 
\label{eq:pkSZmod1}
\end{equation}
where $b_{\rm cl}$ is the average galaxy cluster bias, $\xi^{\delta \delta} (r,z)$ is the scale and redshift dependent  linear density--density correlation function (i.e. the Fourier transform of the linear matter density power spectrum), and $\xi^{\delta \,v}(r,z)$ refers to the density -- distance projected peculiar velocity correlation function, given by
\begin{equation}
\xi^{\delta\, v} (r) \equiv \langle \delta (\vx)\, \hat{\vrv}\cdot \vv(\vx+\vrv)  \rangle = 
 -\frac{H(z)f(z)}{2\pi^2(1+z)}\int dk\,k P_m(k,z)\,j_1(kr),
\label{eq:dvx}
\end{equation}
with $H(z)$ the Hubble parameter, $f(z)=d\log{D}/d\log a$ is the logarithmic derivative of the density linear growth factor with respect to the cosmological scale factor, and $P_m(k,z)$ is the scale and redshift dependent matter power spectrum.

We use the simulated maps of the baseline \CORE-150 mission in the different frequency bands to compute an estimate of the signal-to-noise (S/N) ratio for the detection of the kSZE pairwise momentum for each of those bands. For this, we compute an estimate of the kSZE pairwise momentum in 100 {\it rotated/displaced} configurations of the positions of the $\sim $52,000 clusters 
on each frequency map. That is, we first consider the angular positions of our simulated cluster set and rotate them in galactic longitude on 100 {\it rotated/displaced} configurations. This procedure preserves the real relative angular distances among galaxy clusters. For each rotated configuration, aperture photometry (AP) is conducted on the displaced position of each cluster, with an aperture that equals the maximum between $5\,$arcmin and the FWHM of each map. This is motivated by the fact that the average angular virial radius of our cluster sample is close to $5\,$arcmin.

Each rotated configuration of our cluster sample provides an estimate of the kSZE pairwise momentum on angular positions where we expect to find no kSZE signal. Thus the ensemble of the 100 $p_{kSZ}(r)$ estimates can be used to estimate a covariance matrix for its measurement,
\begin{equation}
\covC_{i,j} = \langle (p_{kSZ}(r_i) - \langle p_{kSZ}(r_i) \rangle) (p_{kSZ}(r_j) - \langle p_{kSZ}(r_j) \rangle \rangle,
\label{eq:cov1}
\end{equation}
where the $\langle ... \rangle$ denote averages of the 100 rotated/displaced configurations. With this covariance matrix, we next adopt a matched filter approach in which the observed data would be fit to our fiducial expectation provided by Eq.~\ref{eq:pkSZmod1}, yielding an amplitude $A$:
\begin{equation}
A = \frac{{\mathbf p_{kSZ,\,obs}}^t \covC^{-1} {\mathbf p_{kSZ,\,{\rm fid}} } }{{\mathbf p_{kSZ,\,{\rm fid}}}^t \covC^{-1} {\mathbf p_{kSZ,\,{\rm fid}} } }.
\label{eq:Amp}
\end{equation}
In this equation, ${\mathbf p_{kSZ,\,obs}}$ refers to the observed kSZE pairwise momentum in the form of a vector evaluated over a set of cluster pair separations $r$, while ${\mathbf p_{kSZ,\,{\rm fid}} }$ corresponds to the fiducial model of Eq.~\ref{eq:pkSZmod1}. 
The formal variance of the amplitude $A$ is given by
\begin{equation}
\sigma^2_A = \frac{1 }{{\mathbf p_{kSZ,\,{\rm fid}}}^t \covC^{-1} {\mathbf p_{kSZ,\,{\rm fid}} } }.
\label{eq:sgAmp}
\end{equation}
If we also assume that our kSZE pairwise momentum is unbiased, then $\langle {\mathbf p_{kSZ,\,obs}}\rangle = {\mathbf p_{kSZ,\,{\rm fid}} }$ and the S/N is given by
\begin{equation}
{\rm S/N} \equiv \frac{\langle A\rangle}{\sigma_A}= \sqrt{{\mathbf p_{kSZ,\,{\rm fid}}}^t \covC^{-1} {\mathbf p_{kSZ,\,{\rm fid}} } }.
\label{eq:sgAmp}
\end{equation}

\begin{figure}[tbp]
\centering
\includegraphics[width=.45\textwidth]{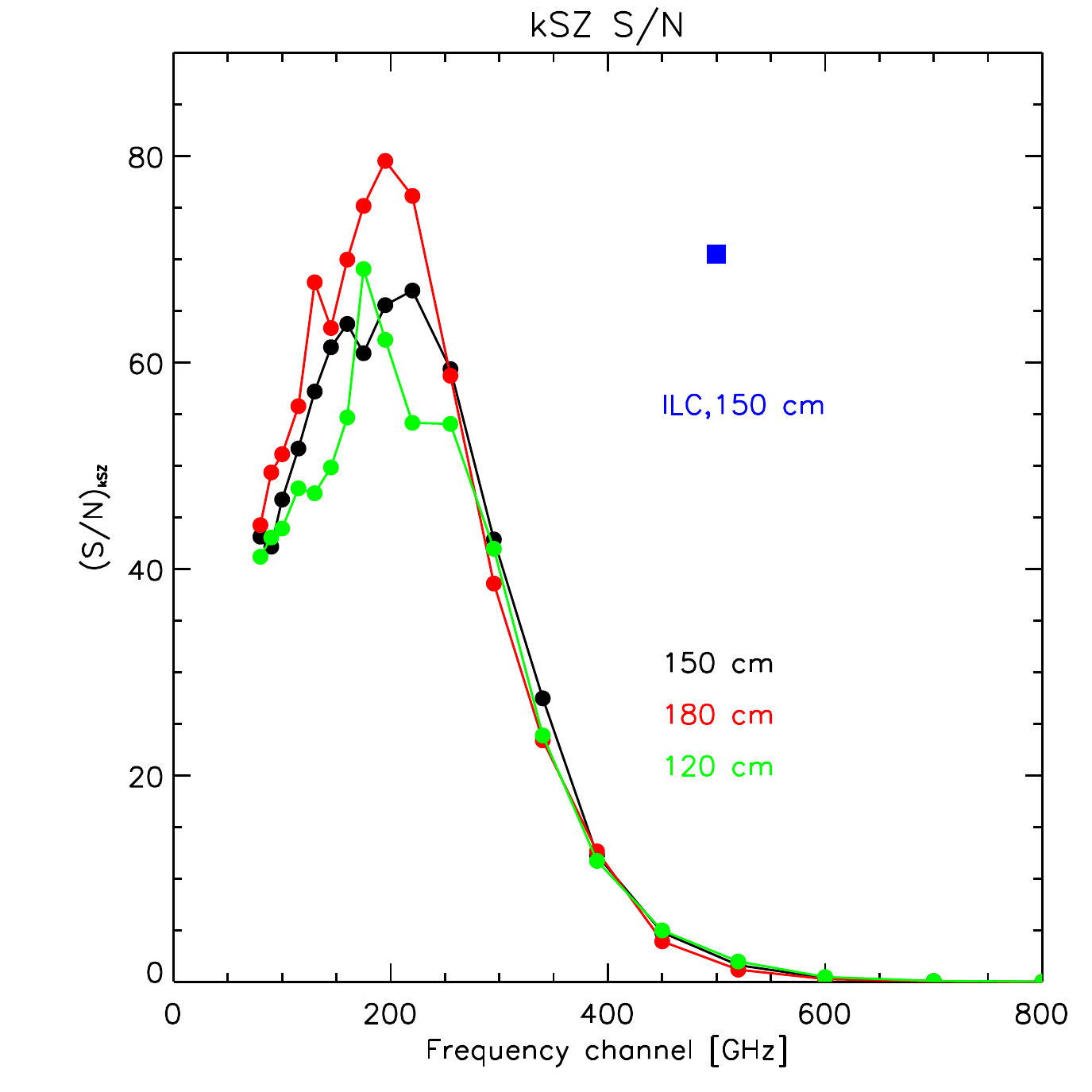} \includegraphics[width=.5\textwidth]{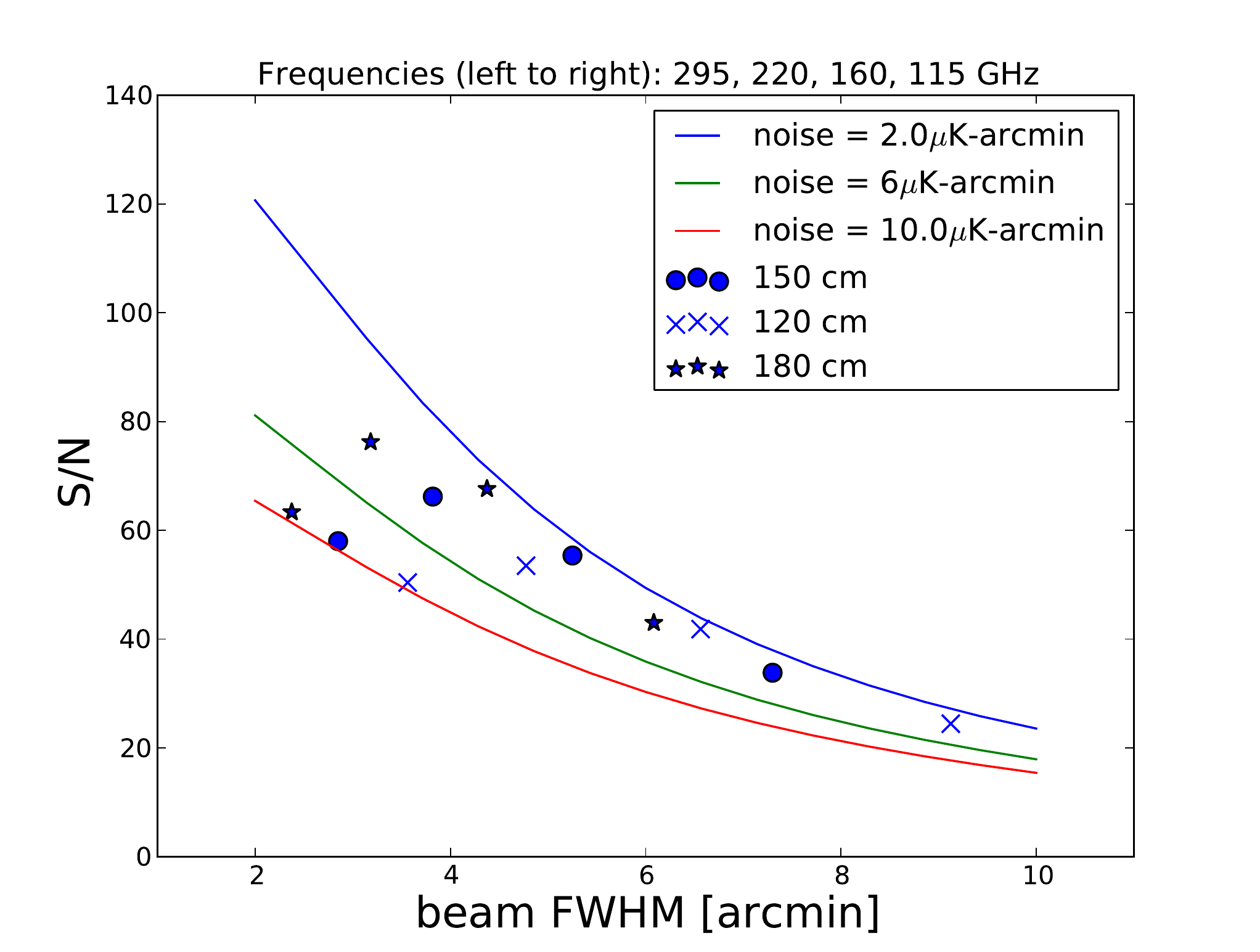}
\caption[fig:S2N1]{{\it Left:} S/N forecasts for the kSZE pairwise momentum obtained from the \corep\ simulated maps. We consider three mirror sizes for all frequency channels, plus a linear combination of all channels exactly projecting the tSZE out (ILC, given by the blue square). {\it Right:} S/N forecasts from a toy model adopting the nominal values for instrumental noise and beam size.}
\label{fig:S2N1}
\end{figure}

In Figure~\ref{fig:S2N1} (left), we plot the S/N obtained for each of the \CORE\ frequency maps, where the three \CORE\ configurations with diameters of 120, 150 and 180~cm are color coded with green, black and red, respectively. 
In the low frequency channels the S/N increases with frequency because the beam FWHM is falling and the CMB contamination of kSZE estimates decreases with the aperture size \citep[see Figure~6 of ][]{chmetal06a}. This trend is however reversed at frequencies close to and above $\nu = 220$\,GHz, because at those frequencies the presence of dust emission in our simulated maps becomes dominant.  We have found that this dust emission contaminating kSZE measurements is predominantly generated within our galaxy.  In the 120 and 180\,cm diameter cases the FWHM of the channels are scaled with respect to the nominal case and this explains why for the low frequency channels the 180\,cm (120\,cm) diameter case is yielding slightly higher (lower) estimated S/N.  At high frequencies ($\nu > 300\,$GHz), the FWHM is typically smaller than 5\,arcmin and the three mirror sizes should yield identical results. However, small-scale dust emission leaks more easily inside the AP filter for larger mirror sizes, thus slightly decreasing the S/N in these cases. The S/N for a linear combination of the different \CORE\ channels that projects out exactly the non-relativistic thermal Sunyaev-Zel'dovich component~\citep{remazeilles2011} is given by the blue square, for the nominal case of 150\,cm diameter size. Because the channels achieving the highest S/N are dominated by CMB residuals, the gain of combining different channels is modest, reaching a maximum of S/N$\simeq 70$.

The behavior of these S/N estimates obtained from realistic frequency maps simulated for the \corep\ mission can be checked against a simple modeling of the kSZE angular power spectrum which can be made dependent on the number of clusters available, their average mass, the quality of the velocity reconstruction, the FWHM or beam size of the CMB experiment, and the instrumental noise level. The predictions for this versatile S/N estimation approach is given in Figure~\ref{fig:S2N1} right, which is in rough agreement with the more realistic estimates of Figure~\ref{fig:S2N1} left.

The kSZE pairwise momentum is dependent on the Thomson optical depth ($\tau_T$), the Hubble parameter and the velocity growth factor $\sigma_8 f(z)$ (see Eq.~\ref{eq:pkSZmod1} above). Using the $p_{kSZ}(r)$ covariance matrix inferred from the simulated maps, it is also possible to provide Fisher-matrix forecasts for the sensitivity of \CORE\ to the factor $\sigma_8f(z)$ after assuming some prior knowledge on the gas content  of clusters ($\tau_T$) and the Hubble parameter.  In fact, a Gaussian prior on the error equal to 5\,\% of the fiducial values of those two parameters yields a typical error in $\sigma_8 f(z)$ at the level of 7\,\% for the entire cluster set. This level of uncertainty is comparable to (and slightly better than) that current $\sigma_8 f(z)$ measurements from experiments like BOSS or WiggleZ \citep{bossfsg,wigglezfsg}. Constraints on this parameter from \Euclid\ will be typically provided for higher redshifts ($z>0.8$), while other experiments like J-PAS \citep[][]{jpas} and SKA \citep[e.g. ][]{abdalla} are expected to be significantly more sensitive (at the percent and sub-percent level), and split into multiple redshift shells. However, further kSZE-based constraints on $\sigma_8 f(z)$ can be obtained from smaller and more abundant halos (of $M\sim 10^{12-13}\,$M$_{\odot}$) which should shrink further kSZE derived $\sigma_8 f(z)$ uncertainties, and whose contribution could also be split in different redshift intervals. Even in a case where kSZE derived $\sigma_8 f(z)$ measurements may not be the most precise, they will likely contribute to shrinking the uncertainties in this factor while suffering from different systematics, and thus providing robustness to the joint estimates, \citep[see][for more detailed forecasts]{SpergelkSZforecasts}.

\subsection{CMB Temperature Evolution with Redshift}
\label{sec:cmb_temp}

If the expansion of the Universe is adiabatic and the CMB spectrum was a black-body at the time it originated, this shape will be preserved with its temperature evolving as
\begin{equation}
T(z) = T_0(1+z)
\end{equation}
This is a robust prediction of standard cosmology, but there are many non-standard, but nevertheless theoretically well-motivated physical processes in which photon number is not conserved. Examples include a non-perfectly transparent Universe, decaying vacuum cosmologies with photon injection mechanisms, models in which the fine-structure constant varies \cite{Martins:2014iaa}, string theory motivated scenarios where photons mix with other particles such as axions \cite{Jaeckel:2010ni}, many modified gravity scenarios, and so on. We note that in such models the distance-duality---or Etherington \cite{Etherington33}---relation is also violated \cite{Avgoustidis:2011aa}, and that mechanisms that lead to deviations from the standard evolution would, in general, create spectral distortions~\citep[see e.g.][]{chluba2014}.

At a purely phenomenological level, deviations from the standard law have often been parametrized by \cite{Lima:1995kd}
\begin{equation}
T(z) = T_0(1 + z)^{1-\beta}\,.
\end{equation}

Various measurements of $T(z)$ already exist, but the currently large uncertainties do not allow for strong constraints on the underlying models to be set. At low redshifts, the $T(z)$ relation can be measured via the tSZE towards galaxy clusters. This method was first applied to ground-based CMB observations \cite{Battistelli:2002ie,Luzzi:2009ae}, which demonstrated its potential. With a new generation of ground based SZE experiments and the all-sky \Planck\ SZE survey, it became possible to use this method to place tight constraints on the redshift evolution of the CMB temperature \cite{Hurier:2013ona,saro14,deMartino:2015ema,Luzzi:2015via}.

At higher redshifts (typically $z>1.5$), $T(z)$ can be evaluated from the analysis of quasar absorption line spectra which show atomic and/or ionic fine structure levels excited by the photon absorption of the CMB radiation. The CMB is an important source of excitation for species with transitions in the sub-millimeter range. Although the suggestion is more than four decades old, measurements (as opposed to upper bounds) were only obtained in the last two decades \cite{Srianand:2000wu}, and the best ones so far still have errors at the ten percent level \cite{Noterdaeme:2010tm}. The current best constraints, combining both direct SZE and spectroscopic measurements as well as indirect distance duality measurements, are\footnote{Constraints on $\beta$ can also be obtained using primary CMB only at $z \sim 10^3$ (see e.g. \cite[Section 6.7.3 of][]{planckcosmo2016} and~\cite{coreparam2016}).} \cite{Avgoustidis:2015xhk}:
\begin{equation}
\beta = (7.6 \pm 8.0)\times 10^{-3}\,.
\end{equation}

Spectroscopic measurements of $T(z)$ using CO are currently signal-to-noise limited and will significantly improve with ELT-HIRES. Individual measurements of the temperature are expected to reach $\Delta T < 0.1$ K, which corresponds to $\sigma_\beta = 8 \times 10^{-3}$. This single-measurement constraint further improves as the square-root of the number of measurements at $z\sim 2 - 3$, of which one can reasonably expect approximately 10. Overall we thus expect ELT-HIRES to deliver a constraint of $\sigma_\beta = 2.5 \times 10^{-3}$.

The possibility of determining $T_{\rm CMB}(z)$ from measurements of the SZE was suggested long ago \cite{fabbri1978,reph1980}.   Compton scattering of the CMB in a cluster causes an intensity change, $\Delta I_{\rm SZ}$, that can be written as:
\begin{equation}
\Delta I_{SZ}={
2 k^{3} T_0^{3} \over h^{2}c^{2} } {x^4e^x \over (e^x -1)^2} 
\tau \biggr[\theta f(x) - \beta + R(x, \theta, \beta) \biggl
] \, ,
\end{equation}
where $\tau =\sigma_T \int n_e dl $ is the optical depth, $T_0$ is the CMB temperature at redshift $z=0$, 
$\theta=\frac{k_BT_e}{m_ec^2}$ where $T_e$ is the cluster electron temperature (we are assuming isothermality), $\beta= {v_z}/{c}$ 
where $v_z$ is the radial component 
of the peculiar velocity of the cluster, $f(x)=[x\coth(\frac{x}{2})-4] $, and the $R(x,\theta,\beta)$ function 
includes relativistic corrections \cite{yoel1995,itoh2002,Shimon2002}.

The spectral signature of the SZE, $\Delta I_{\rm SZ}$, depends on the 
frequency $\nu$ through the dimensionless frequency $x=\frac{h\nu(z)}{kT(z)}=\frac{h\nu_0}{kT_0}$: it is 
redshift-invariant only for the standard scaling of $T_{\rm CMB}(z)$. In all other non-standard scenarios,
the `almost' universal dependence of SZE on frequency becomes $z$-dependent,
resulting in a small dilation/contraction of the SZE spectrum on the frequency axis. 
In terms of thermodynamic temperature, $\Delta T_{\rm SZ}$ of the CMB due to the SZE is given by  
\begin{equation}
	\Delta T_{\rm SZ}(x) = T_0 \tau [\theta f(x) -\beta   + R(x,\theta,\beta)]  
\end {equation}
If we assume that $T_{\rm CMB}$ scales with $z$ as $T_{\rm CMB}(z)=T_0(1+z)^{1-\beta}$, while the frequency scales as
$(1+z)$ as usual, then the dimensionless frequency will be $x^{\prime}=\frac{h\nu_0}{k_B T^{\ast}_{\rm CMB}}$ and
$T^{\ast}_{\rm CMB}=T_{\rm CMB}(z)/(1+z)=T_0(1+z)^{-\beta}$ will be slightly
different from the local temperature $T_0$ as measured
by \COBE-FIRAS. In this way it is possible to measure the temperature of
the CMB at the redshift of the cluster, thus directly 
constraining scenarios such as those discussed above.  
Actually, what we measure is the temperature change integrated over the solid angle corresponding to the 
chosen aperture radius $\theta_1={\rm max}[\theta_{500}, 0.75\times\theta_{FWHM}(\nu)]$ (as defined in~\cite{Luzzi:2015via}), so we have:
\begin{equation}
	\Delta \mathfrak{T}_{\rm SZ} = \int \Delta T_{\rm SZ}(x) d\Omega = T_0 [\theta f(x) -\beta   + R(x,\theta,\beta)]\int\tau d\Omega~~. 
\end {equation}
In the following we will use $\mathcal{T}_{500}$ the integral of $\tau$ in a sphere of radius $R_{500}=\theta_{500} \times D_A(z)$ with $D_A(z)$ the angular diameter distance of the cluster.

We have simulated the observations of a cluster sample selected from the \CORE\ cluster catalog according to the following requirements:
\begin{itemize}
\item redshift $z>0.5$, because of the higher lever arm on $\beta$;
\item $\theta_{500}>2.5$ arcmin, to keep under control the level of foregrounds (at this step assumed to be previously removed) in the aperture photometry;
\item $\mathcal{T}_{500} >0.03 \, {\rm arcmin}^2$, this last choice is done mainly to keep high S/N clusters.   
\end{itemize}
This selection provides a sample of 48 high S/N clusters.

The frequency bands considered in the simulation are between 90~GHz and 340~GHz, because the lower and higher frequencies are best suited for foreground extraction and are not 
useful for the reconstruction of the SZE signal. CMB and foregrounds are removed first, adopting the procedure described in~\cite{Luzzi:2015via}. 

The forecast for the SZE signal for the clusters has been obtained from the mock cluster catalog, assuming a generalized NFW pressure model. We have used the \CORE\ noise level (values as reported in Table~\ref{tab:core}) to estimate the errors in the observed spectra. The beam dilution effect has been taken into account to estimate errors on the SZE signal for each channel.

The mock dataset was then analyzed to recover the original input parameters of the cluster. The analysis has been performed through a MCMC algorithm, which allows us to explore the full space of the cluster parameters (integrated optical depth $\tau$, peculiar velocity $v_{pec}$, electron temperature $T_e$) and the CMB brightness temperature at the redshift of the cluster. In the analysis we allow for calibration uncertainties, considered as an uncertain scale factor, which was modeled with a Gaussian distribution with mean 1 and standard deviation $0.1\%$. 
We assume that the maps have been cleaned of foregrounds and CMB. To take into account possible CMB and kinematic SZE residuals after the removal, we model them as a kinematic SZE component and adopt as a prior a Gaussian with zero mean and with a $500 \, {\rm km}{\rm s}^{-1}$ standard deviation. We include a prior on the cluster gas temperature, assuming that these clusters have an X-ray counterpart with electron temperatures known with a $1\sigma$ error of 0.1~keV.  These priors can be obtained with \eROSITA, which can measure cluster temperatures to an accuracy of 10 - 35\%, depending on the cluster temperature (ranging from 1-2 to 10~keV)~\citep{merloni12}. The need for a good knowledge of $T_{\rm e}$ is mainly dictated by the fact that the MCMC converges more quickly with respect to the case in which the prior on $T_{\rm e}$ is broad. We have tried for a small sample of 25 clusters using a broader prior on $T_e$ and the resulting constraints on $T_{\rm CMB}(z)$ are unchanged.

\begin{figure}[tbp]
\centering
\includegraphics[width=.7\textwidth]{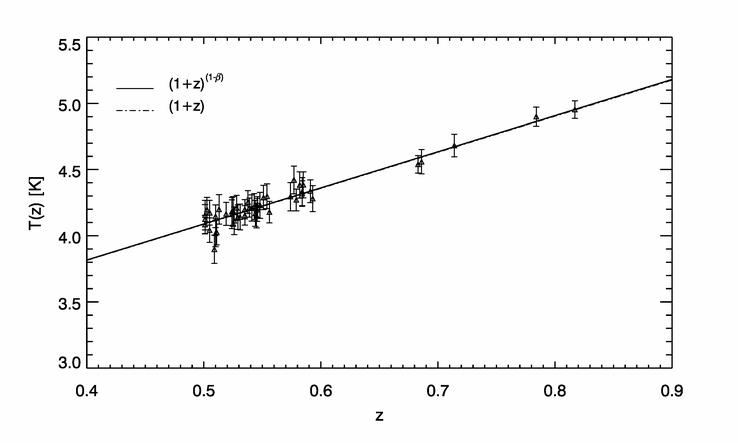}
\caption[fig:tcmb]{Recovered temperatures from a high-S/N subsample of the mock \CORE\ cluster catalogue and the associated fit (solid line). The expected dependence in the standard $\Lambda$CDM model is $1+z$ (dash-dotted line).}
\label{fig:tcmb}
\end{figure}

To obtain $\beta$ we have performed a fit of the $T(z)$ data points as shown in Figure~\ref{fig:tcmb}. The final $\beta$ value we get by fitting the $T(z)$ data points obtained with the MCMC treatment is
\begin{equation}
\beta = -0.0011 \pm 0.0054\,.
\end {equation} 

\CORE\ will do significantly better than \Planck. The study has been restricted to a few tens of clusters for simplicity. With a sample of clusters of a few hundreds, we would end up with an error bar that is more than two times lower ($\sigma_\beta \lesssim 2.7\times 10^{-3}$). There is, of course, room for improvement by enlarging the sample size.  Therefore, \CORE --- in combination with ELT-HIRES --- offers the possibility of mapping the redshift evolution of the CMB temperature at the percent level or better for individual sources all the way from $z=0$ to beyond $z=3$, and to improve current constraints on the parameter $\beta$ by at least one order of magnitude.

\section{Diffuse SZ Emission}
Additional information is available from analysis of the map of the diffuse SZ emission over the sky.  Here we examine how well \CORE\ can extract an SZ map of the sky and how to use the one-dimensional PDF of that map as a cosmological probe.
 
\label{sec:diffuse}
\subsection{SZE Map}
We reconstruct the Compton-$y$ parameter map of the tSZE effect with the Needlet Internal Linear Combination 
\citep[NILC,][]{delabrouille2009, remazeilles2011, Remazeilles2013}.  {\tt NILC} performs a weighted linear combination of the \CORE\ sky frequency maps such that the combination is of minimum variance, while the weights assigned to each frequency map offer unit response to the tSZE frequency spectrum. While the second condition guarantees a reconstruction of the tSZE signal without any bias by projecting the observations onto the tSZE frequency scaling vector, the first condition ensures that the emission from other foreground components and the noise are minimized. By decomposing the \CORE\ sky maps onto a frame of spherical wavelets, called needlets \citep{Narcowich2006}, {\tt NILC} adapts foreground cleaning to the local conditions of contamination both over the sky and over the angular scales, therefore optimizing the component separation. The inverse squared RMS of the global 'noise' (foregrounds + instrumental) residual contamination in the {\tt NILC} $y$-map is the sum of the inverse squared RMS of the noise in each \CORE\ frequency map, such that the {\tt NILC} reconstruction will always benefit from an increased number of frequency channels.

\begin{figure}[t!]
\centering
\includegraphics[width=0.5\textwidth]{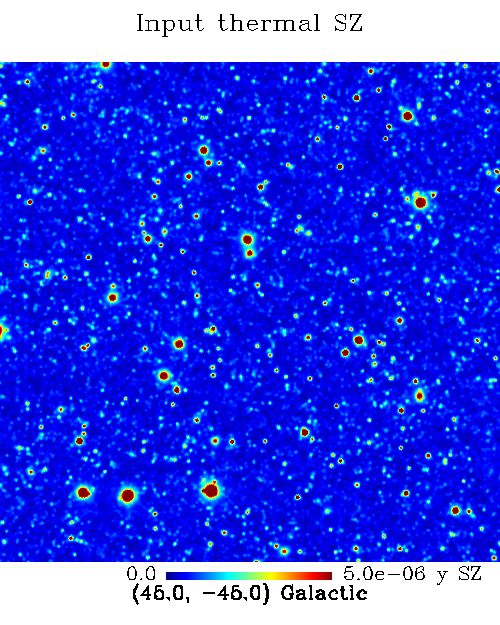}~
\includegraphics[width=0.5\textwidth]{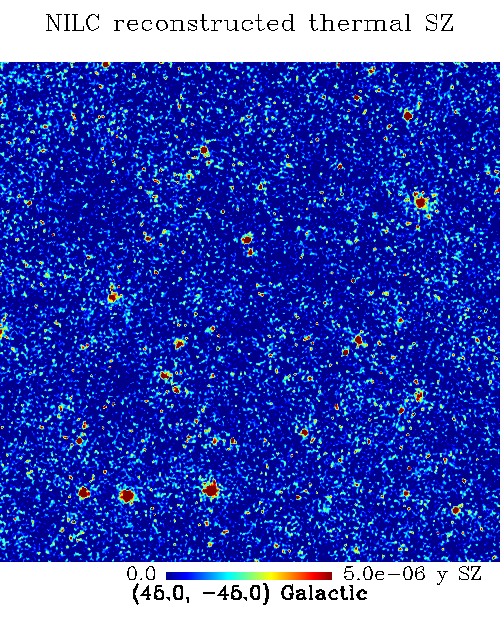}~\\
\caption{\corep\ {\tt NILC} tSZE map versus input tSZE map in a $12^\circ.5 \times 12^\circ.5$ patch of the sky centred at $(\ell,b)=(45^\circ,-45^\circ)$.}
\label{fig:nilc_gn}
\end{figure}

The large number of frequency bands ($19$) of \CORE\ allows us to reconstruct the tSZE emission with unprecedented accuracy on the sky. 
Fig.~\ref{fig:nilc_gn} compares the {\tt NILC} tSZE y-map with the input tSZE y-map of the simulation in a $12^\circ.5 \times 12^\circ.5$ area of the sky centred at $(\ell,b)=(45^\circ,-45^\circ)$. We can see how accurately \CORE\ extracts the tSZE emission from a large number of galaxy clusters, with minimum residual contamination from foregrounds and noise.


\begin{figure}[t!]
\centering
\includegraphics[width=0.8\textwidth]{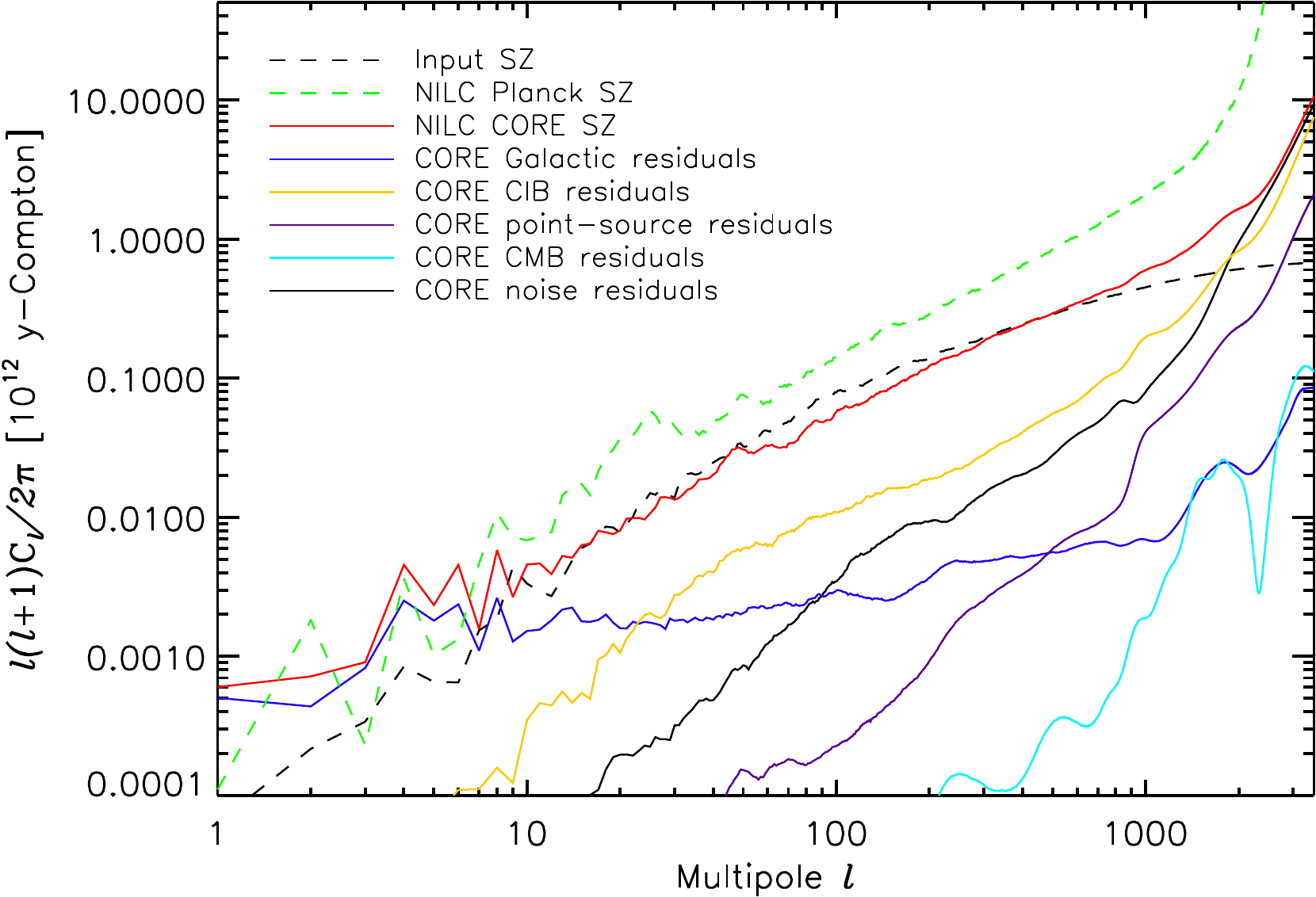}
\caption{Angular power spectrum of the \CORE\ tSZE map (solid red line) with respect to the input tSZE power spectrum (dashed black line), along with the power spectrum of residual foregrounds after component separation by {\tt NILC}. The tSZE power spectrum from a \emph{Planck}-like mission is over-plotted in dashed green line for a side-by-side comparison with the performance of \CORE.}
\label{fig:ps_sz}
\end{figure}

In Fig.~\ref{fig:ps_sz} we show the angular power spectrum of the \CORE\ tSZE map (solid red line) along with the power spectrum of the residual foreground components and residual noise (coloured solid lines) leaking into the \CORE\ tSZE map after component separation with {\tt NILC}. The power spectra of the residual contamination have been computed by applying the {\tt NILC} weights that go in the reconstructed tSZE map to the individual component maps of the simulation. We also plot in Fig.~\ref{fig:ps_sz} the power spectrum of the fiducial tSZE signal (dashed black line). The reconstructed tSZ power spectrum by \CORE\ matches almost perfectly the input tSZE power spectrum on angular scales $10 < \ell < 1000$. Galactic foregrounds, CIB, radio sources, CMB, and instrumental noise are clearly controlled and minimized by more than one order of magnitude with respect to the cosmological tSZE signal. For the sake of comparison, we over-plot as dashed green line the reconstructed tSZE power spectrum by {\tt NILC} for a simulation of the \emph{Planck} mission, in which case residual foreground contamination after component separation clearly dominates the tSZE signal over the whole range of angular scales. While \emph{Planck} requires  marginalizing a posteriori over foreground residuals in the likelihood estimation of $\sigma_8$ from the reconstructed tSZE power spectrum \citep{Planck_2015_XXII}, and therefore is model-dependent, \CORE\ allows us to accurately recover the tSZE signal over a large range of angular scales without the need of a posteriori marginalization over residuals.  \CORE\ enables a fully data-driven reconstruction of the tSZE signal over the sky at high angular resolution.

\subsection{SZ Map Statistics: Estimating $\sigma_8$ with the thermal SZE 1-PDF from \CORE\ simulations}
We also performed an analysis of the 1D probability distribution function (PDF) of the
reconstructed Compton parameter map. For the tSZ effect, we expect an asymmetric 1D PDF
distribution with a significantly positive tail \citep{2003MNRAS.344.1155R}.
Using this fact, we can derive a constraint on the parameter $\sigma_8$ by fitting
that positive tail against the one extracted from simulated Compton $y$-maps.
The simulated PDF was built from the map by binning the values of $y$ after applying
 the $\sim 37\%$ sky mask, obtaining the reference distribution $P(y)$. The theoretical PDF was computed following the 
formalism described in \cite{Hill2014}. We used the same set of cosmological parameters employed for simulating 
the Compton map, and considered the \cite{Tinker2008} mass function. The Compton profile for individual clusters
 was determined using the \cite{arnaud2010} pressure profile, normalized using the scaling relations described in
 \cite{planck2013xx}, and considering the same mass bias parameter employed for the simulation, $b=0.37$.
 The Compton profile was smoothed with a Gaussian kernel to account for the \CORE\ 4 arcmin beam.

We fixed all cosmological parameters except for $\sigma_8$, which we varied to build a likelihood function by fitting the
 theoretical $P(y)$ histogram with the reference one. The cluster $y$-profile introduces correlations between different 
$y$ bins. This was accounted for when computing the likelihood by using a multivariate Gaussian with a covariance 
matrix which was evaluated numerically, accounting also for the pixel to pixel Poisson term. The range of values of $y$
 chosen for the fit was the one dominated by the cluster contribution. We chose to use the interval from $y \simeq 7.3 \times 10^{-6}$ 
 to $y \simeq 37.3 \times 10^{-6}$.  In this way we discarded the very high $y$ region, were noise due to poor statistics becomes a major issue, and the very low $y$  region, where the distribution is dominated by other foregrounds and the approximation of non-overlapping clusters, implicit in our computation of the $P(y)$, is no longer applicable.

We found the final estimate $\sigma_8=0.814\pm0.002$ ($68\%$ C.L.). This value is compatible with the one
 used for generating the simulated map, $\sigma_8=0.815$. In \cite{planck2015xxii} a similar analysis was
 performed to estimate $\sigma_8$ from the Planck all-sky Compton parameter map, yielding the estimate $\sigma_8=0.77\pm 0.02$.  We note that our result reduces the uncertainty by a factor $\sim 10$. We also checked that this estimate
 is robust when choosing a different upper limit for the $y$ interval considered for the fit.

\begin{figure}[t!]
\centering
\includegraphics[width=0.6\textwidth]{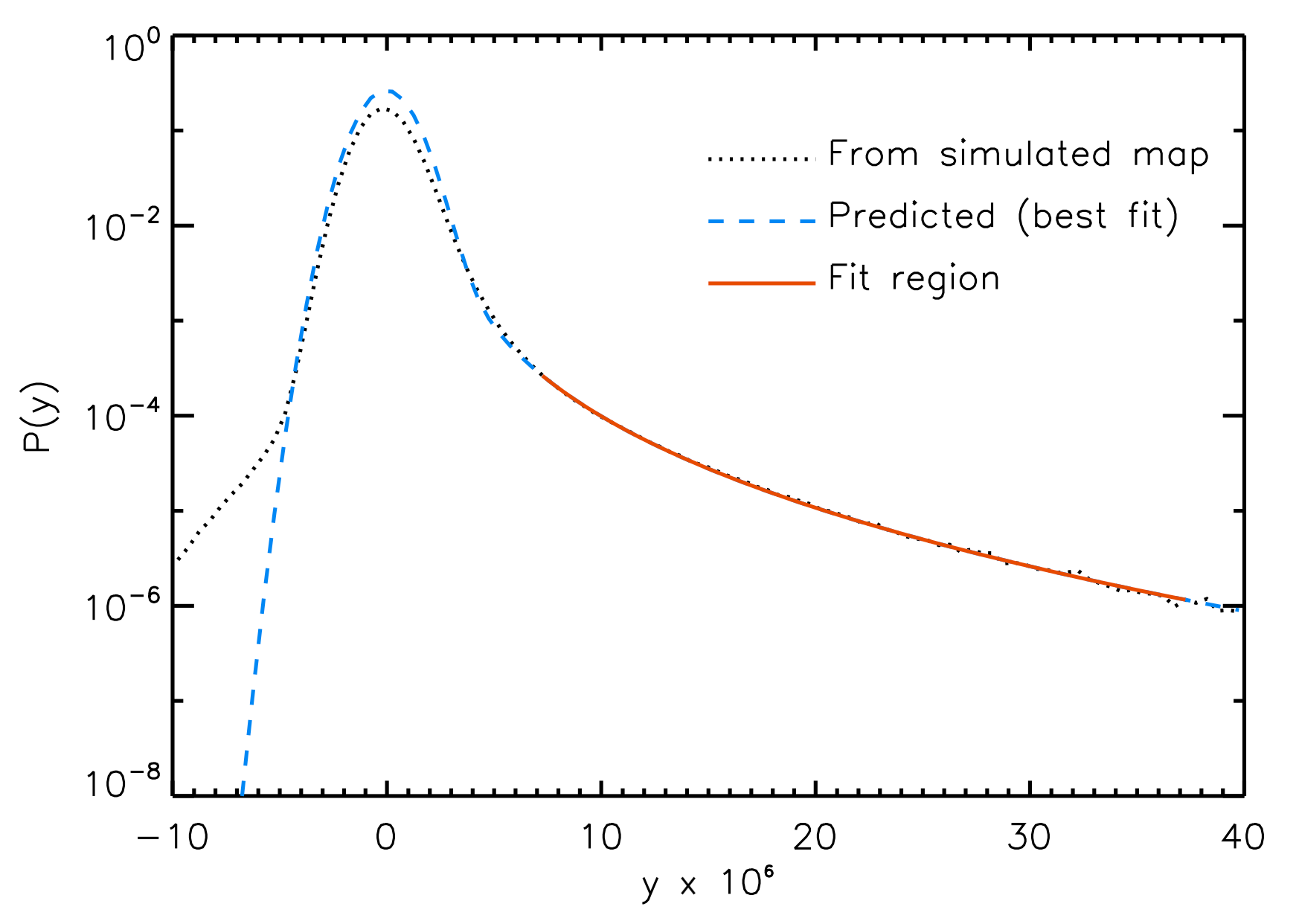}
\caption{Compton parameter 1D-PDF employed for the $\sigma_8$ parameter estimation. Dotted: distribution extracted from the \CORE\ map realization; dashed: best-fit computed PDF, corresponding to our estimate $\sigma_8=0.814$; solid: region considered for the fit. Low values of $y$ were discarded from this analysis because the modelling of the PDF in this region does not accurately account for signal arising from overlapping clusters along the line-of-sight or for sources of contamination affecting the lowest $y$ bins.
\label{fig:Py}
}
\end{figure}

\section{Conclusions}
\label{sec:conclusions}

We have examined in detail the capabilities of the \CORE\ mission for galaxy cluster science, exploring the impact of primary mirror size, and comparing \CORE\ to options for a future ground-based CMB-S4 experiment.  \CORE-150 (1.5\,m primary) will detect $\sim 50,000$ clusters in its all-sky survey ($\sim 15\%$ sky mask).  The yield increases (decreases) by $25\%$ if one increases (decreases) the telescope primary mirror size by 30~cm.  An observatory with 3.5\,m telescopes and 155,000 detectors observing in five frequency bands from the Atacama plateau, CMB-S4 (Atacama) would be shallower than \CORE, while a South Pole observatory observing with a 10\,m telescope instead, CMB-S4 (South Pole), would be significantly deeper. The combination of catalogues from the two ground sites (Atacama+South Pole) would provide $\sim 80,000$ clusters. In combination with CMB-S4 (South Pole), \CORE-150 would reach a limiting mass of $M_{500} \sim 2-3 \times 10^{13} M_\odot$ and detect well over 200,000 clusters, including $\sim 20,000$ clusters at $z>1.5$ if we assume that the gas in these young structures follows scaling laws similar to the local scaling laws with self-similar evolution in redshift -- an open question that such future CMB observations will allow us to investigate in some depth.

In the base $\Lambda$CDM framework, \CORE\ cluster counts alone will provide competitive constraints on $\Omega_{\rm m}$ and $\sigma_8$ when the other parameters are constrained by, e.g., primary CMB and/or non-CMB external datasets.  Full exploitation of the catalogue, however, requires the cluster mass scale to be determined to better than a few percent (<5\%) accuracy (Sec.~\ref{sec:sig8-omega}); this will be achieved over a broad redshift range using CMB halo lensing stacks. \CORE\ cluster counts will also be competitive for constraining the dark energy equation of state parameter $w_0$ and its possible redshift evolution $w_a$ in the $w\Lambda$CDM model when marginalizing over all other parameters ($\sigma_{w_0}=0.28$, $\sigma_{w_a}=0.31$).  Adding cluster counts to \CORE\ primary CMB will tighten the constraints on the dark energy equation--of--state parameters ($\sigma_{w_0}=0.05$, $\sigma_{w_a}=0.13$), and improve the ability to measure neutrino mass, reaching an uncertainty of $\Sigma m_\nu = 39 \, {\rm meV}$ ($1\sigma$).

\CORE\ will measure cluster masses using CMB halo lensing out to the highest redshifts, well beyond the range of galaxy shear observations.  Individual masses can be measured with uncertainties of  $4\times10^{14}/2\times10^{14}/ 10^{14} \, M_\odot$ (at $1\sigma$) using CMB halo lensing for \CORE-120/\CORE-150/\CORE-180, respectively.  A large telescope ($\gsim 180 \, {\rm cm}$) would be required to determine individual masses with uncertainties at the level $\lsim 10^{14} \, M_\odot$. Stacking the CMB lensing signal on many clusters will calibrate cluster scaling relations to a few percent out to redshift $z=1.5$ for \CORE-120, and out to $z=2$ for \CORE-150 and \CORE-180.  This component of the \CORE\ science is crucial for constraining cosmology with cluster counts.

\CORE\ will use clusters in a number of additional studies.  It will provide extremely precise and accurate measurements of the composite SZE spectrum for combinations of ten or more clusters.  With this information one can study the effects of the relativistic thermal SZE as well as contaminants like intracluster dust.   For the highest mass clusters, it will be possible to measure ICM temperatures using the relativistic effect, although cluster temperature measurements may well be hampered by emission from the intracluster dust that contaminates the measured flux at the higher frequencies.  The broad frequency coverage afforded by \CORE\ enables this experiment.  Pairwise momentum through the kSZE can be extracted to $S/N>60$ in individual frequency channels between 150 and 300~GHz; at higher frequencies, dust from our Galaxy decreases the S/N.  Finally, the tSZE can be used as a precision test of one of the basic tenants of the standard model, the redshift evolution of the CMB temperature.  Deviation from the standard evolution of $T_{\rm CMB}(z) \propto (1+z)$ will be constrained to better than $5.4 \times 10^{-3}$.

The size of the \CORE\ primary telescope mirror only weakly impacts detection of the relativistic tSZE, the kSZE pairwise momentum significance and the measurement of $T_{CMB}(z)$, but wide frequency coverage is needed to disentangle the various signals (primary CMB, Galactic emission and cluster contaminants such as intracluster dust emission).  Cosmological parameters from cluster counts also depend only weakly on telescope size, provided the cluster mass scale is determined to a few percent.  This will be enabled through a CMB halo lensing analysis using the \CORE\ dataset; however, this measurement benefits from a larger aperture.  An aperture $>150 \, {\rm cm}$  is needed to determine the cluster mass scale to a few percent out to redshift $z=2$, and to measure individual masses with uncertainties of the order of $10^{14} M_\odot$. The $>150 \, {\rm cm}$ aperture would also enable detection of hundreds of clusters at $z>1.5$ and push the redshift limit of the survey well beyond the limits of the \eROSITA\ and \Euclid\ surveys.

Finally, we have shown that the broad frequency coverage of \CORE\ enables clean extraction of a full-sky map of diffuse SZE signal, much improved over the \Planck\ SZE map by significantly reducing contamination from foregrounds.  Our analysis of the one-dimensional distribution of Compton-$y$ values in the simulated map finds an order of magnitude improvement in the possible constraints on $\sigma_8$ over a similar analysis carried out on the \Planck\ data, demonstrating that this will be a powerful cosmological probe.

\acknowledgments

Some of the results in this paper have been derived using the HEALPix~\citep{gorski2005} package. Parts of the cosmological analysis was made using Cosmo MC~\citep{lewis2002}, CLASS~\citep{blas2011} and MontePython~\citep{audren2012}. C.H.-M. acknowledges financial support of the Spanish Ministry of Economy and Competitiveness via I+D project AYA-2015-66211-C2-2-P. CJM is supported by an FCT Research Professorship, contract reference IF/00064/2012, funded by FCT/MCTES (Portugal) and POPH/FSE. J.G.N. acknowledges financial support from the Spanish MINECO for a 'Ramon y Cajal' fellowship (RYC-2013-13256) and the I+D 2015 project AYA2015-65887-P (MINECO/FEDER).
\bibliographystyle{JHEP}
\bibliography{coreclusters}

\end{document}